\documentclass[10pt]{article}

\usepackage[table,xcdraw]{xcolor}   % loads »colortbl«
\usepackage{amsmath,amssymb,amsthm, graphicx, mathrsfs, pgf,tikz,marvosym,fontawesome,slashed, mathtools,array,bbm,yfonts, adjustbox, adforn, fourier-orns, relsize,stmaryrd,simplewick,youngtab,ytableau,youngtab, epigraph, url, bclogo,arydshln,multirow,dsfont, lipsum, tensor,feynmf, wrapfig, enumitem,verbatim}
\usepackage{pgfplots} \pgfplotsset{compat=1.15}

\usepackage{physics}
\usepackage[left=20mm,right=20mm,top=35mm,bottom=35mm,columnsep=15pt]{geometry} 
\usepackage{placeins,multirow}
\usepackage[T1]{fontenc}
\usepackage[utf8]{inputenc}
\usepackage[nottoc]{tocbibind}
\usetikzlibrary{arrows}
\usetikzlibrary{positioning}
\usepackage[english]{babel}
\usepackage[caption=false]{subfig}

\makeatletter
\let\@fnsymbol\@arabic
\makeatother

\author{T. Hertog\thanks{thomas.hertog@kuleuven.be}, S. Maenaut\thanks{simon.maenaut@kuleuven.be}, B. Missoni\thanks{bruno.missoni@gmail.com}, R. Tielemans\thanks{rob.tielemans@skynet.be}, T. Van Riet\thanks{thomas.vanriet@kuleuven.be}}
\title{\textbf{Stability of Axion-Saxion wormholes}}
\date{}

\usepackage{titlesec}
\usepackage{cancel}
\usepackage{tikz-cd}

\definecolor{r}{cmyk}{1,.50,0,.20} 

\usepackage{array,mathtools,amssymb,booktabs}
\newcolumntype{C}{>{$}c<{$}}

\usepackage[subfigure]{tocloft}

\usepackage[square,comma,numbers]{natbib}

\usepackage[colorlinks=true, linkcolor=r, urlcolor=r, linktocpage=true, citecolor=r]{hyperref}

\numberwithin{equation}{section}

\newcommand*\dif{\mathop{}\!\mathrm{d}}
\newcommand{\diff}{\mathrm{d}}
\newcommand{\A}{\mathrm{A}}
\newcommand{\B}{\mathrm{B}}
\newcommand{\C}{\mathrm{C}}
\newcommand{\D}{\mathrm{D}}
\newcommand{\E}{\mathrm{E}}
\newcommand{\F}{\mathrm{F}}
\newcommand{\G}{\mathrm{G}}
\newcommand{\HH}{\mathrm{H}}
\newcommand{\I}{\mathrm{I}}
\newcommand{\J}{\mathrm{J}}
\newcommand{\M}{\mathrm{M}}
\newcommand{\PP}{\mathrm{P}}
\newcommand{\R}{R}
\newcommand{\GS}{\Phi}

\newcommand{\dg}[1]{\delta \tensor{g}{#1}}

\newcommand{\dx}[1]{\tensor{\dd{x}}{#1}}
\newcommand{\dF}[1]{\delta \tensor{F}{#1}}
\newcommand{\dch}{\delta \tensor{\chi}{}}

\newcommand{\ph}{\varphi}
\newcommand{\Chi}{\mathcal{X}}
\newcommand{\dph}{\delta \tensor{\ph}{}}

\newcommand{\dpsi}{\tensor{\psi}{}}
\newcommand{\dyg}{\mathcal{Y}}
\newcommand{\dA}{\tensor{A}{}}
\newcommand{\dB}{\tensor{B}{}}
\newcommand{\dE}{\tensor{E}{}}
\newcommand{\Ppsi}{\Pi_{\psi}}
\newcommand{\Pdph}{\Pi_{\dph}}
\newcommand{\Pdch}{\Pi_{\dch}}

\newcommand{\Py}{\Pi_{\dyg}}
\newcommand{\Pf}{\Pi_{f}}
\newcommand{\PE}{\Pi_{E}}
\newcommand{\PR}{\Pi_{R}}
\newcommand{\PPh}{\Pi_{\Phi}}
\newcommand{\PCh}{\Pi_{\Chi}}
\newcommand{\Rs}{\mathcal{R}}
\newcommand{\Rt}[1]{\tensor{\mathcal{R}}{#1}}

\newcommand{\Ls}[2][]{\tensor*[#1]{\mathcal{L}}{#2}}

\newcommand{\Hh}{\mathcal{H}}
\newcommand{\kh}{k}
\newcommand{\dchd}{\delta \tensor{\Dot{\chi}}{}}
\newcommand{\chid}{\Dot{\chi}}
\newcommand{\Hhd}{\Dot{\mathcal{H}}}
\newcommand{\phd}{\Dot{\varphi}}

\newcommand{\dphd}{\delta \tensor{\Dot{\ph}}{}}

\newcommand{\dpsid}{\tensor{\Dot{\psi}}{}}

\newcommand{\dEd}{\tensor{\Dot{E}}{}}
\newcommand{\kapd}{}
\newcommand{\kapf}{}
\newcommand{\kaps}{}
\newcommand{\kape}{}

\newcommand{\gt}[1]{\tensor{g}{#1}}
\newcommand{\gm}[1]{\tensor{\gamma}{#1}}

\newcommand{\xv}[1]{\tensor{x}{#1}}
\newcommand{\Ft}[1]{\tensor{F}{#1}}
\newcommand{\xt}[1]{\tensor{\xi}{#1}}
\newcommand{\xtd}[1]{\tensor{\Dot{\xi}}{#1}}
\newcommand{\CD}[1]{\tensor{\nabla}{#1}}
\newcommand{\cd}[1]{\tensor{D}{#1}}

\newcommand{\eps}[1]{\tensor{\epsilon}{#1}}
\newcommand{\Eps}[1]{\tensor{\mathcal{E}}{#1}}

\begin{document}%

\maketitle

\begin{center}
\emph{Institute for Theoretical Physics, KU Leuven,\\Celestijnenlaan 200D, 3001 Leuven, Belgium} 
\end{center}

\begin{abstract}
We reconsider the perturbative stability of Euclidean axion wormholes. The quadratic action that governs linear perturbations is derived directly in Euclidean gravity. We demonstrate explicitly that a stability analysis in which one treats the axion as a normal two-form gauge field is equivalent to one performed in the Hodge-dual formulation, where one considers the axion as a scalar with a wrong-sign kinetic term. Both analyses indicate that axion wormholes are perturbatively stable, even in the presence of a massless dilaton, or saxion, field that couples to the axion. 
%At no point does our quadratic action suffer from the conformal factor problem.
\end{abstract}

\newpage
\tableofcontents

\newpage

\section{Introduction}
%The saddle point expansion in Euclidean gravity has been shown to be a powerful tool, yet it remains incomplete since it ignores the UV completion of Einstein gravity. String theory provides such a UV completion and with AdS/CFT we even have access to a full non-perturbative definition of quantum gravity in backgrounds with certain asymptotics. This presents an opportunity to understand the semi-classical saddle point expansion; whether and how one should include various topologies.  This is where 

For many years, the study of Euclidean wormhole solutions in semi-classical gravity theories has proven useful to gain some understanding of certain non-perturbative properties of quantum gravity.

The so-called Giddings-Strominger wormholes \cite{Giddings:1987cg}, where the wormhole throat is supported by axion-flux, are perhaps the prime examples of Euclidean wormhole solutions \cite{Giddings:1988cx, Giddings:1989bq, Coleman:1988cy, Lavrelashvili:1987jg}\footnote{For a review of the vast literature on this subject see \cite{Hebecker:2018ofv}, and \cite{Martucci:2024trp} for very recent applications.}. In recent years, more general solutions were found in bottom-up models, including wormholes sourced by multiple fields such as massive dilatons \cite{Andriolo:2022rxc, Jonas:2023ipa, Jonas:2023qle}, and higher-derivative corrections \cite{Andriolo:2020lul, Cheong:2023hrj}. In parallel, axion wormholes and variations thereof were embedded in string theory and in the AdS/CFT setting \cite{Loges:2023ypl, Hertog:2017owm, Astesiano:2022qba,Marolf:2021kjc, Astesiano:2023iql, Anabalon:2023kcp}\footnote{See \cite{Hamada:2019fmc} for a compendium on holographic backgrounds sourced by axion densities.}. These embeddings sharpened the paradoxes that such wormhole-saddles of a gravitational path integral appear to induce. These paradoxes range from the factorisation problem \cite{Maldacena:2004rf} and the absence of Coleman's $\alpha$-parameters in AdS/CFT duals \cite{Arkani-Hamed:2007cpn}, to certain Swampland principles \cite{McNamara:2020uza} and a violation of operator positivity in dual CFTs \cite{Katmadas:2018ksp, Loges:2023ypl}. Quite clearly, a resolution of these paradoxes will require a better understanding of the wormhole solutions, in particular under what conditions they are stable.

Euclidean axion wormholes either connect two different universes, or different regions within the same universe. Here we concentrate on rotationally-invariant solutions in flat space, which is, just like anti-de Sitter (AdS) space, a setting where wormholes connect different universes. This is in contrast with Euclidean de Sitter (dS) space, where wormholes connect opposite sides of the same Euclidean sphere \cite{Gutperle:2002km,Aguilar-Gutierrez:2023ril}, yielding a kettle-bell-like geometry. In this case, wormhole configurations can contribute to the Hartle-Hawking quantum state \cite{Aguilar-Gutierrez:2023ril}. 

Building on recent work \cite{Hertog:2018kbz,Loges:2022nuw}, we study the perturbative stability of axion wormholes in the presence of a dilaton field, or saxion. As argued long ago by Coleman \cite{Coleman:1987rm}, the existence (and number) of negative modes around saddle points in Euclidean quantum gravity determines their nature and interpretation. If there are no negative modes, one expects the saddle point to lift a vacuum degeneracy or contribute to the wave function of the universe. Saddles and boundary conditions for which there is a single negative mode describe tunnelling transitions. Finally, the presence of a multitude of negative modes indicates that the saddle point in question may not provide a physically meaningful or relevant contribution to the gravitational path integral. 

Originally Coleman \cite{Coleman:1988cy} suggested that axion wormholes describe tunnelling events in which baby universes pinch of from or are absorbed by the mother universe. This would lead to a violation of axion charge conservation from the viewpoint of the mother universe, consistent with the expectation that gravity eliminates global symmetries through wormholes \cite{Kallosh:1995hi}. From this perspective wormholes contribute to matrix elements described by path integrals with boundary conditions that keep the axion charge fixed. 
%A possible resolution of this lies in the boundary conditions. If the axion charge is not fixed, the homogenous mode becomes dynamical, as explained in \cite{Hertog:2017owm}, and then it was shown it leads to a negative mode \cite{Rubakov:1996cn}.  This is similar to cosmological instantons where the tunnelling negative mode is removed by fixing boundary conditions. 
In this context, \cite{Hertog:2018kbz, VanRiet:2020csu} argued that axion wormholes have multiple negative modes, and this was interpreted as evidence that e.g. the factorisation problem may not be a problem after all.  However, subsequently \cite{Loges:2022nuw} studied the stability of these wormholes based on a different treatment of the axion perturbations, and found no negative modes. Whereas \cite{Hertog:2018kbz} worked with the scalar field formulation of the axion, the analysis of \cite{Loges:2022nuw} made use of the Hodge-dual formulation, where the axion is a two-form field. Since Hodge duality frames cannot change physical results, here we reconsider the analysis of \cite{Hertog:2018kbz}. We identify a problem with the boundary condition used in \cite{Hertog:2018kbz} and show in detail how the boundary condition that the axion charge remains fixed, when implemented correctly, resolves the discrepancy between the two Hodge dual formulations in \cite{Hertog:2018kbz} and \cite{Loges:2022nuw}. Along the way we generalise the stability analysis to axion wormhole saddles that include a massless dilaton field, since this is expected on general grounds in string theory embeddings of these solutions \cite{Hertog:2017owm, Astesiano:2022qba, Loges:2023ypl}.
%although dilaton masses can appear \cite{Loges:2023ypl} as well.  We also differ with previous works by directly perturbing Euclidean gravity instead of Wick rotating known results in cosmological perturbation theory \cite{Gratton:1999ya, Gratton:2000fj}. 

{\bf Note added:} While this paper was in the final stages of completion, \cite{Jonas:2023qle} appeared which verified the stability of the homogeneous sector. Here we consider all modes.

\section{Axion-Saxion wormholes}

We consider wormhole solutions of Euclidean gravity coupled to an axion field. An axion $\chi$ is a scalar which enjoys a classical shift symmetry $\chi\rightarrow \chi+c$. It can be dualised to a two-form field $B_{\mu\nu}$ in four dimensions with field strength $F=dB$ related to $\chi$ via $\star_4 F_3=\pm d\chi$. In Euclidean signature Hodge duality can be subtle, however, since axions can come with a `wrong-sign' kinetic term, depending on what boundary conditions one considers. In what follows we are interested in boundary conditions that keep the axion charge $Q$ or better, the axion flux, fixed:
\begin{equation}
    Q =\frac{1}{2\pi^2} \int_\Sigma F. \label{eq: charge axion 2}
\end{equation}
where $\Sigma$ is a hypersurface at infinity. 

The metric of wormhole solutions that preserve rotational symmetry can be written in the form
\begin{align}\label{metric}
    \dif s^2 = N^2(r)\dif r^2 + a^2(r)\dif\Omega_{3}^2, 
\end{align}
where $\dif\Omega_{3}^2$ is the metric on $S^{3}$. The field strength is then of the form
\begin{equation}
    F = Q\mathcal{E},
\end{equation}
where $\mathcal{E}$ is the volume form on $S^3$. The equations of motion can be deduced from the following action:
\begin{equation}
    S[B, g] = \int\left[-\frac{1}{2\kappa_4^2}\star\mathcal{R}+\frac{1}{2}\star dB\wedge dB\right] \label{eq: action 3 formv1}
\end{equation}
The $B$-field equation is solved by our Ansatz above and the Einstein equation reduces to 
\begin{equation}\label{Friedmann}
    \left(\frac{\dot{a}}{N}\right)^2 = 1 - \frac{\kappa_4^2Q^2}{6} a^{-4},
\end{equation}
where $\dot a = \partial_{r} a$. In conformal gauge $(N=a)$ the solution reads $a^2=\frac{\kappa_4|Q|}{\sqrt{6}}\cosh(2r)$.  The inclusion of a cosmological constant in \eqref{eq: action 3 formv1} is straightforward, see e.g. \cite{Gutperle:2002km}. 

To exhibit the choice of boundary conditions clearly at the level of the action, it is useful to write the action as a function of $F$, 
\begin{equation}
    S[F, \chi, g] = \int\left[-\frac{1}{2\kappa_4^2}\star \mathcal{R}+\frac{1}{2}\star F\wedge F + \chi \dif F\right] . \label{eq: action 3 formv2}
\end{equation}
Here $\chi$ is a Lagrange multiplier that enforces $dF=0$ such that, locally, $F=dB$. This action clearly reproduces the same equations of motion, but has the benefit that it allows a clean implementation of the above boundary condition as a Dirichlet condition on $F$. Alternatively, one can integrate out $F$ and leave it as a function of $\chi$, which is the proper way to carry out Hodge dualisation at the level of the action. This yields
\begin{equation}
    F = -\star\dif \chi,\label{eq: Hodge}
\end{equation}
which, when inserted in the action gives
\begin{equation}
    S[\chi, g] = \int_{\mathcal{M}}\left[-\frac{1}{2\kappa_4^2}\mathcal{R}-\frac{1}{2}\star\dif\chi\wedge\dif\chi\right] -\int_{\partial\mathcal{M}}\dif\left(\chi \star\dif\chi\right). \label{eq: action scalar}
\end{equation}
The $\chi$-field has the wrong-sign kinetic term. Yet, the action is well-defined because of the boundary term, which renders the non-gravitational part bounded from below. 
In this Hodge-dual form, the Noether charge of the shift symmetry gives the axion flux
$Q = (2\pi^2)^{-1}\int_\Sigma\dif^3x\sqrt{\gamma}\:\pi$, where  $\pi=\partial^0\chi$ is the Euclidean momentum of $\chi$.
   
In the context of unified theories such as string theory axions tend to be part of a complex scalar. The partner field $\varphi$, the dilaton or saxion, does not enjoy a shift symmetry, but typically has a Lagrangian of the form
\begin{equation}
\mathcal{L} =- \frac{1}{2}\star d\varphi \wedge d\varphi +  \frac{1}{2}e^{b\varphi}\star d\chi \wedge d\chi   + d(e^{b\varphi}\chi\star d\chi) . \label{axio-dilaton}
\end{equation}
If the dilaton-coupling $b$ is sufficiently small,%{\color{red}TH: inconsistent notation: you use $\kappa$ here and at the start of the next section, and $\kappa_4$ elsewhere in this section}
\begin{equation}\label{regularity}
    b^2 < \frac{8}{3}\kappa_4^2\ ,
\end{equation}
then this axion-dilaton theory has regular wormhole solutions \cite{Arkani-Hamed:2007cpn}. An example is type II string theory on a six-torus, which reduces to maximal supergravity in 4 dimensions. This admits a consistent truncation of the moduli space such that one obtains the above axion-dilaton model with a coupling $b$ in the range \eqref{regularity}  \cite{Bergshoeff:2004pg, Loges:2023ypl}. This shows that regular axion-saxion wormhole solutions can be embedded in string theory.

In general, supersymmetric compactifications of string theory lead to general sigma models coupled to gravity with Lagrangian
\begin{equation}
    \mathcal{L} =- \frac{1}{2} G_{IJ}(\Phi)\star d\Phi^I \wedge d\Phi^J + \ldots
\end{equation}
where $G_{IJ}$ is the moduli metric and the dots represent total derivatives. Once again there are symmetric instanton solutions of the form \eqref{metric}. In these, the scalars trace out geodesics in the target space with metric $G_{IJ}$, parametrized by an affine coordinate $h(r)$ \cite{Breitenlohner:1987dg}. The latter is nothing but the solution for the radial harmonic on \eqref{metric}:
\begin{equation}
\partial_{r}(a^3 N^{-1}\partial_r h(r)) =0 \quad \rightarrow \quad \frac{d h}{dr} =  N a^{-3}\,.
\end{equation}
The total geodesic velocity is a constant
\begin{equation}
G_{IJ}(\Phi)\frac{d}{d h}\Phi^I\frac{d}{d h}\Phi^J = c \,,    
\end{equation}
where $c$ must be negative in order to have a regular wormhole. In effect, using this geodesic behaviour one recovers the single-axion equation \eqref{Friedmann} for the scale factor with $Q^2 = -c$.
% replaced by $-c$ in \eqref{Friedmann}.\footnote{The single axion model is reproduced since then $\chi=Qh$.}
%When the sigma model has commuting Killing vectors, they correspond to axion directions and then there is an adapted coordinate frame on the target space that turns the symmetries into shift symmetries and the corresponding directions $\chi^i$ are the axions that can be dualised to 2-forms. In the scalar duality frame this then generates the wrong sign kinetic terms and the required total derivatives for Euclidean actions in which one keeps the charges fixed. 
For the two-field model \ref{axio-dilaton}, the explicit solutions for the geodesics are:
\begin{equation}
 e^{b\varphi/2} = \frac{|Q|}{\sqrt{-c}}\cos\left(\frac{b\sqrt{-c}}{2} \,h(r)\right)\,,\qquad \chi= \frac{2\sqrt{-c}}{bQ}\tan\left(\frac{b\sqrt{-c}}{2} \,h(r)\right) \,.  \label{geo} 
\end{equation}
The regularity condition \eqref{regularity} follows from the requirement that $e^{b\varphi/2}$ should not change sign as $h(r)$ varies over the entire wormhole background\footnote{This requirement might seem rather ad hoc. Note, however, that on general grounds one interprets the dilaton coupling to $(\partial\chi)^2$ as the squared coupling constant of the axion two-form $B$. This is confirmed explicitly in the top-down constructions of axion wormholes in flat space \cite{Loges:2023ypl} where $e^{b\varphi/2}$ represents the physical size of certain cycles inside the compactification space and hence are required to remain positive.}. In conformal gauge the harmonic function reads (up to an additive constant),
\begin{equation}
   h(r) = \frac{\sqrt{6}}{2\kappa_4\sqrt{-c}}\arctan\left(\sinh(2r)\right) \ ,
\end{equation}
which, when substituted in \eqref{geo}, yields the regularity condition \eqref{regularity}.
% is now apparent since the expression for $e^{b\varphi/2}$ contains a cosine function with argument $b\sqrt{-c}h(r)/2$ which takes values over an interval of size $\frac{b\sqrt{6}}{4\kappa_4}\pi$. Demanding that the cosine function does not flip sign within this interval restricts the size of the interval to $\pi$ and we reproduce the condition \eqref{regularity}. 
\newline

\section{Perturbations}\label{stabilitysec}

\subsection{Perturbed action}

To study whether axionic wormholes can be relevant saddle point contributions to the gravitational path integral, we now analyse the stability of the solutions in the two-field model with respect to small fluctuations. To this end we expand the action up to quadratic order in perturbations in both the metric and the (s)axion fields. 
%The result will be independent of the formulation of the axion field fluctuations if the correct boundary conditions are imposed. 
We set $\kappa_4=1$ from now on and consider first the action in terms of the three-form field strength, viz.
\begin{align}
\label{eqn:action-field-strength-three-form}
S_E &= \int \dd[4]{x} \sqrt{g} \left( - \frac{1}{2} \Rs + \frac{1}{12} e^{-b \ph} \Ft{_{\mu\nu\sigma}} \Ft{^{\mu\nu\sigma}} + \frac{1}{2} \CD{_\mu} \ph \CD{^{\mu}} \ph + \frac{1}{6} \chi \eps{^{\mu\nu\rho\sigma}} \CD{_\mu} \Ft{_{\nu\rho\sigma}} \right).
\end{align}
As discussed above, the Lagrange multiplier in \eqref{eqn:action-field-strength-three-form} ensures that this is equivalent to the action in the scalar field formulation with the appropriate total derivative added,
\begin{align}
\label{eqn:action-field-scalar}
S_E &= \int \dd[4]{x} \sqrt{g} \left( - \frac{1}{2} \Rs - \frac{1}{2} e^{b \ph} \CD{_\mu} \chi \CD{^\mu} \chi + \frac{1}{2} \CD{_\mu} \ph \CD{^\mu} \ph + \CD{_\mu} \left( e^{b\ph} \chi \CD{^\mu} \chi \right) \right).
\end{align}
In this section we perform a stability analysis of the wormholes in both the three-form and scalar-field formulations of the theory. We demonstrate explicitly that the results of both calculations agree, provided boundary conditions are treated carefully. 

\begin{comment}
We write the fluctuations of the metric and the fields in a four-dimensional covariant manner:
\begin{align}
\gt{_{\mu\nu}} &\to \gt{_{\mu\nu}} + \dg{_{\mu\nu}},&
\Ft{_{\mu\nu\rho}} &\to \Ft{_{\mu\nu\rho}} + \dF{_{\mu\nu\rho}} &
\chi &\to \chi + \dch& \ph &\to \ph + \dph.
\end{align}
Indices are raised and lowered by the metric $\gt{_{\mu\nu}}$ and perturbations act on the field strength with lowered indices.

To study how fluctuations transform under a change of coordinates (diffeomorphism), we consider an infinitesimal gauge transformation of the form $\xv{^\mu} \to \xv{^\mu} - \xt{^\mu}$ under which the fluctuations transform as \cite{Mukhanov:1992me}
\begin{align}
\dg{_{\mu\nu}} &\to \dg{_{\mu\nu}} + \Ls{_\xi} \gt{_{\mu\nu}},&
\dF{_{\mu\nu\rho}} &\to \dF{_{\mu\nu\rho}} + \Ls{_\xi} \Ft{_{\mu\nu\rho}} &
\dch &\to \dch + \Ls{_\xi} \chi& \dph &\to \dph + \Ls{_\xi} \ph.
\end{align}
where $\Ls{_\xi}$ is the Lie derivative along $\xi$. To avoid any possible issues with gauge fixing, it will be crucial to carry out the wormhole stability analysis with gauge invariant variables that satisfy the boundary conditions. 
\end{comment}

The background equations of motion are given by
\begin{align}\label{eqn:equation-of-motion}
\Hh^2&= \kh + \frac{1}{6} \left( \phd^2 - e^{b \ph} \chid^2 \right)  = \kh + \frac{1}{6}\left(\phd^2 - e^{-b\ph} Q^2/a^4 \right),&
\Dot{\Hh} &= - 2 \Hh^2 + 2 \kh,\\
\ddot{\ph} &= - 2 \, \Hh  \, \phd - \frac{1}{2}  \, b  \, e^{b \ph} \, \chid^2 = - 2  \, \Hh  \, \phd - \frac{1}{2}  \, b  \, e^{-b \ph} \, Q^2 / a^4, &
\ddot{\chi} &= - 2 \Hh \chid - b \phd \chid.
\end{align}
where $\Hh$ is the conformal Hubble rate $\dot{a}/a$ and we have kept for now the three-space curvature $k$ general.

The theory of perturbations around a wormhole background of the form \eqref{metric} is analogous to cosmological perturbation theory in Euclidean signature (see e.g. \cite{Mukhanov:1992me,Gratton:1999ya}). There and here, it is useful to decompose a general metric perturbation in so-called scalar, vector and tensor components, which to linear order evolve independently. Since the vector and tensor components aren't sourced by scalar fields, they aren't potential sources of instability. Thus we concentrate here on scalar metric perturbations. In conformal gauge, the wormhole metric perturbed by a general scalar perturbation reads 
\begin{align}\label{eqn:scalar-metric-perturbation}
\dd{s}^2 &= \left(\gt{_{\mu\nu}} + \dg{_{\mu\nu}}\right) \dx{^\mu} \dx{^\nu}\notag\\
&=a^2 \left[ \left( 1 + 2 \dA \right) \dd{r}^2 + 2 \cd{_i} \dB \dd{r} \dd{x}^i + \left( \left( 1 - 2 \dpsi \right) \gm{_{ij}} + 2 \cd{_i} \cd{_j} \dE \right)\dx{^i} \dx{^j}  \right],
\end{align}
where $\gm{_{ij}}$ and $\cd{_i}$ are the induced metric and the covariant derivative on the unit three-sphere. Latin indices are raised and lowered with the induced metric $\gm{_{ij}}$.  In addition, we have the axion and dilaton fluctuations $\dch$ and $\dph$.

%Scalar perturbations are only sensitive to gauge transformation of the form $\xt{^\mu} = \left(\xt{^0},\cd{^i} \xi\right)$ where $\xt{^0}$ and $\xi$ are scalar functions \cite{Mukhanov:1992me}. Scalar metric perturbations then transform as
%\begin{align}
%\dA &\to \dA + \Hh \xt{^0} + \xtd{^0},&
%\dB &\to \dB + \xt{^0} + \xtd{},&
%\dpsi &\to \dpsi - \Hh \xt{^0}&
%\dE &\to \dE + \xi 
%\end{align}
%where a dot represents a derivative in the $r$-direction. 
%The perturbations of the scalar fields $\ph$ and $\chi$ transform as
%\begin{align}
%\dph &\to \dph + \xt{^0} \phd, & \dch \to \dch + \xt{^0} \Dot{\chi}.
%\end{align}

In the three-form formalism, we parametrise axion fluctuations as in \cite{Loges:2022nuw}
\begin{align}\label{eqn:perturbed-three-form}
\dF{} = \frac{1}{6} f \Eps{_{ijk}} \dx{^i} \wedge \dx{^j} \wedge \dx{^k} + \frac{1}{2} \Eps{_{ijk}} \cd{^k} w \dd{r} \wedge \dx{^i} \wedge \dx{^j}
\end{align}
where $\Eps{_{ijk}}$ is the volume form of the unit three-sphere. Since $\Ft{}$ is closed this implies that
\begin{equation}\label{dotf=deltaw}
 \Dot{f} = \triangle{w},
\end{equation} 
where $\triangle{} = \cd{_i} \cd{^i}$ is the Laplacian on the unit three-sphere. Under gauge transformations $\xv{^\mu} \to \xv{^\mu} - \xt{^\mu}$, with $\xt{^\mu} = \left(\xt{^0},\cd{^i} \xi\right)$, these fluctuations transform as
\begin{align}
f &\to f + Q \triangle{\xi},& w &\to w + Q \xtd{}.
\end{align}

The perturbative stability of wormholes depends on the quadratic action operator and on the Dirichlet boundary conditions on the axion charge and dilaton fluctuation. Since all physically meaningful statements are gauge-invariant, we demand the same from our boundary conditions. Perturbations of the axion charge and of the dilaton field transform under gauge transformations as
\begin{align}\label{gi boundary conditions}
 \delta Q_{\text{axion}} &\to \delta Q_{\text{axion}} + \int_{S^3} \dd[3]{x} \sqrt{\gm{}} Q \triangle{\xi}& \dph &\to \dph + \xt{^0} \phd.
\end{align}
The integral over the three-sphere vanishes since its integrand is a total derivative and the three-sphere is a compact manifold without a boundary. The term with $\phd$ also vanishes since $\ph$ asymptotically approaches a constant value. Thus, our boundary conditions are gauge-invariant and we can safely proceed with our analysis.

% 3.1
% (Einstein Hilbert term) details in appendix

We expand the action to quadratic order in the perturbations using the Mathematica package $x$Pand \cite{Pitrou:2013hga}, which
uses the tensor algebra package $x$Tensor in the $x$Act distribution \cite{Martin:2008xtensor} together with the package $x$Pert,
for perturbations \cite{Brizuela:2008ra}.\footnote{For this the $x$Pand package was modified to calculate
conformal perturbations in Euclidean instead of Lorentzian spacetime \cite{Maenaut:2020Git}.}
The perturbed Euclidean Einstein-Hilbert action reads,
%is perturbed to second order, where, up to a total derivative, all the remaining terms are the product of two first-order perturbations that form the quadratic action given by
\begin{comment}
\begin{align}
\delta^2 S_E^\text{gr.} = - &\frac{1}{2} \int \dd[4]{x} \sqrt{\gt{}} \Bigg[ \frac{1}{4} \left( \CD{_\mu} \dg{^\nu_\nu} \right) \left( \CD{^\mu} \dg{^\lambda_\lambda} \right)
 - \frac{1}{2} \left( \CD{^\mu} \dg{^\lambda_\lambda} \right) \left( \CD{_\nu} \dg{_\mu^\nu}\right) + \frac{1}{2} \left( \CD{_\mu} \dg{_{\lambda\nu}} \right) \left( \CD{^\nu} \dg{^{\lambda\mu}} \right)
  \notag\\ 
&- \frac{1}{4} \left( \CD{_\nu} \dg{_{\lambda\mu}} \right) \left( \CD{^\nu} \dg{^{\lambda\mu}} \right) + \dg{_\alpha^\nu} \dg{^{\alpha\mu }} \Rt{_{\mu\nu}}
 - \frac{1}{2} \dg{^\alpha_\alpha} \dg{^{\mu\nu}} \Rt{_{\mu\nu}}
 - \frac{1}{4} \dg{_{\alpha\mu}} \dg{^{\alpha\mu}} \Rs
 + \frac{1}{8} \dg{^\alpha_\alpha} \dg{^\mu_\mu} \Rs \Bigg] .
\end{align}
Note that second-order perturbations of the metric and the fields are disregarded as they are multiplied with background expressions in the perturbed action that are zero due to the equations of motion. Selecting the scalar perturbations, we fill in the metric fluctuation from \eqref{eqn:scalar-metric-perturbation}, which yield, 
\end{comment}
up to a total derivative in $r$ and a covariant total derivative on the induced metric, 
%the following quadratic action
\begin{align}
\delta^2 S_E^\text{gr.} =& - \frac{1}{2} \int \dd{r} \dd[3]{x} \Big( a^2 \sqrt{\gamma} \Big[
  \Big(
   6 \left( \dpsid \right)^2
 + 12\Hh \left( \dA + \dpsi \right) \dpsid
 + 9 \Hh^2 \left( \dA + \dpsi \right)^2
 + 4 \Hh \left( \dA + \dpsi \right) \triangle \left( \dB - \dEd \right)
 - 4 \Hh \dpsid \triangle \dE \notag\\
&+ 4 \dpsid \triangle \left( \dB - \dEd \right)
 + 4 \Hh \cd{_i} \dpsi \cd{^i} \dB
 - 6 \Hh^2 \left( \dA + \dpsi \right) \triangle \dE 
 + 4 \Hh \triangle \dE \triangle \left( \dB - \dEd \right)
 - 4 \Hh \triangle E \triangle B
 - 3 \Hh^2 \left( \triangle E \right)^2 \Big)\notag\\
&+ 3 \Hh^2 \cd{_i} \dB \cd{^i} \dB
 - 2 \cd{^i} \dpsi \cd{_i} \left( 2 \dA - \dpsi \right)
 - \kh \Big(
 + 3 ( \dA + \dpsi )^2
 + 2 \, (\dB - \dEd) \triangle (\dB - \dEd)
 - (6 \dA - 2\dpsi + \triangle \dE)  \triangle \dE\notag\\
&+ 3 \,  (\cd{_i} \dB)(\cd{^i} \dB)
 + 4 \,  ( \Hhd+\Hh^2) (\cd{_i} \dE)(\cd{^i} \dE)
 + 2 (\cd{_j} \cd{_i} \dE)(\cd{^j} \cd{^i} \dE)
  \Big)
\Big]\Big).
\end{align}
%where $k$ is linked to the curvature of the induced three-dimensional metric. 
For $k=0$, this agrees with \cite{Mukhanov:1992me}, after Wick rotation and up to the above-mentioned total derivatives.

% 3.2
% (axion + dilaton) both the scalar and three form approach

% dilaton sector perutbation

\begin{comment}
For the field sector of the action, we start with the second-order perturbation of the dilaton term, which is simply a free scalar field. We perturb the metric and the dilaton field, which yields the following quadratic action
\begin{align}
\delta^2 S_E^\text{dilaton} = \int \dd[4]{x} \sqrt{\gt{}} \bigg(&
   \left( -\frac{1}{8} \dg{_{\beta\gamma}} \dg{^{\beta\gamma}}
 + \frac{1}{16} \dg{^\beta_\beta} \dg{^\gamma_\gamma} \right)
\left( \CD{_\alpha} \ph \CD{^\alpha} \ph \right)
 + \frac{1}{2} \dg{^\beta_\beta} \CD{_\alpha} \dph \CD{^\alpha} \ph
 + \frac{1}{2} \CD{_\alpha} \dph \CD{^\alpha} \dph \notag\\
&- \dg{_{\alpha\beta}} \CD{^\alpha} \ph \CD{^\beta} \dph
 + \left( \frac{1}{2} \dg{_\alpha^\gamma} \dg{_{\beta\gamma}}
 - \frac{1}{4} \dg{_{\alpha\beta}} \dg{^\gamma_\gamma} \right)
   \left( \CD{^\alpha} \ph \CD{^\beta} \ph \right) \bigg) .
\end{align}
%
If we select the scalar perturbations of the metric fluctuation \eqref{eqn:scalar-metric-perturbation}, this yields, 
\end{comment}
The perturbed dilaton-field sector of the action is given by, again up to a total derivative in $r$ and a total derivative on the induced metric, 
%the following quadratic action
\begin{align}
\delta^2 S_E^\text{dilaton} = \int& \dd{r} \dd[3]{x} \,  a^2 \sqrt{\gamma}  \, \bigg[
   \frac{1}{2} \left( \dphd^2  + \cd{_i} \dph \cd{^i} \dph \right)
 - \phd \left( \dphd \left( \dA + 3 \dpsi - \triangle \dE \right)
 + \cd{_i} \dph \cd{^i} \dB \right) \notag \\
&+ \frac{1}{4} \phd^2 \left( 3 \dA^2 + 3 \dpsi^2 + 6\dA \dpsi
 - 2 \dpsi \triangle \dE -2 \dA \triangle \dE
 + \cd{_i} \dB \cd{^i} \dB
 + \triangle \dE \triangle \dE
 - 2 \cd{_i} \cd{_j} \dE \cd{^i} \cd{^j} \dE \right)
 \bigg] .
\end{align}

%\newpage

For the axion field, it is useful to demonstrate explicitly that the perturbed action derived from \eqref{eqn:action-field-strength-three-form} is equivalent to the one derived from \eqref{eqn:action-field-scalar}. Starting with the former, the perturbed action up to quadratic order in the perturbations is given by
\begin{align}\label{eqn:perturbed-three-form-field-strength-axion-action}
\delta^2 S_E^\text{axion,F} &= \int \dd[4]{x} \sqrt{g} \bigg[ e^{-b \ph} \bigg( 
  \frac{1}{12} \dF{^{\mu\nu\rho}} \dF{_{\mu\nu\rho}}
 + \frac{1}{12} \dg{^\alpha_\alpha} \Ft{^{\mu\nu\rho}} \dF{_{\mu\nu\rho}}
 - \frac{1}{2} \dg{^{\alpha\beta}} \Ft{^{\mu\nu}_\alpha} \dF{_{\mu\nu\beta}}
 + \frac{1}{4} \dg{^{\alpha\beta}} \dg{^{\kappa\lambda}} \Ft{^{\mu}_{\alpha\kappa}} \Ft{_{\mu\beta\lambda}} \notag\\
&+ \frac{1}{4} \left( \dg{^{\alpha\sigma}} \dg{_\sigma^\beta} - \frac{1}{2} \dg{^\sigma_\sigma} \dg{^{\alpha\beta}} \right)
   \Ft{^{\mu\nu}_\alpha} \dF{_{\mu\nu\beta}}
 + \frac{1}{48} \left( \frac{1}{2} \dg{^\alpha_\alpha} \dg{^\beta_\beta} - \dg{^{\alpha\beta}} \dg{_{\alpha\beta}} \right)
   \Ft{^{\mu\nu\rho}} \Ft{_{\mu\nu\rho}}
   + \frac{1}{24} b^2 \dph^2 \Ft{^{\mu\nu\rho}} \Ft{_{\mu\nu\rho}} \notag\\
&- b \dph \left( 
   \frac{1}{6} \Ft{^{\gamma\mu\nu}} \dF{_{\gamma\mu\nu}}
 + \frac{1}{24} \dg{^\gamma_\gamma} \Ft{^{\mu\nu\alpha}} \Ft{_{\mu\nu\alpha}}
 - \frac{1}{4} \dg{^{\gamma\mu}} \Ft{_\gamma^{\nu\alpha}} \Ft{_{\mu\nu\alpha}}  
  \right) \bigg)
 + \frac{1}{6} \eps{^{\alpha\mu\nu\rho}} \dch \CD{_\alpha} \dF{_{\mu\nu\rho}}
\bigg]
\end{align}
As in \eqref{eq: Hodge}, we can integrate the last term by parts and apply the equations of motion for the field strength fluctuation to find a Hodge-dual relation, expanded to first order in the perturbation: %{\color{red}TH: don't you want this equation referred to with a single number ? I think that you refer to this equation below as 3.13 whereas you mean both 3.13 and 3.14}
%(indices of $F$ have to be lowered when perturbing)
\begin{equation}\label{eqn:perturbed-hodge-duality}
\begin{split}
\Ft{_{\mu\nu\rho}} &= e^{b \ph} \eps{_{\sigma\mu\nu\rho}} \CD{^\sigma} \chi, \\
\dF{_{\mu\nu\rho}} &= e^{b \ph} \eps{_{\sigma\mu\nu\rho}} \left( b\,\dph \CD{^\sigma} \chi + \CD{^\sigma} \dch + \frac{1}{2} \dg{^\alpha_\alpha} \CD{^\sigma} \chi - \dg{^{\sigma\beta}} \CD{_\beta} \chi \right).
\end{split}
\end{equation}
% where we note that the indices of the field strength have to be lowered when applying a perturbation.
% Notice that the first order expression contains the same information as the equation of motion for $\dF{}$ derived from the action above. 

Starting with the scalar-field formulation of the axion action \eqref{eqn:action-field-scalar}, the second-order perturbed action reads
%ation and keep the first-order perturbations of  the metric and the fields, the quadratic action equals
\begin{align}\label{eqn:perturbed-scalar-field-axion-action}
\delta^2 S_E^\text{axion,$\chi$} =  \int \dd[4]{x} \sqrt{\gt{}} \bigg[& e^{b\ph} \bigg(
   \left( \frac{1}{4} \dg{_{\alpha\beta}} \dg{^\gamma_\gamma}
 - \frac{1}{2} \dg{_\alpha^\gamma} \dg{_{\beta\gamma}} \right)
   \left( \CD{^\alpha} \chi \CD{^\beta} \chi \right) 
 + \left( \frac{1}{8} \dg{_{\beta\gamma}} \dg{^{\beta\gamma}}
 - \frac{1}{16} \dg{^\beta_\beta} \dg{^\gamma_\gamma} \right)
   \left( \CD{_\alpha} \chi \CD{^\alpha} \chi \right)\notag\\
&- \frac{1}{2} \CD{_\alpha} \dch \CD{^\alpha} \dch 
 - \frac{1}{2} \dg{^\beta_\beta} \CD{_\alpha} \dch \CD{^\alpha} \chi
 + \dg{_{\alpha\beta}} \CD{^\alpha} \chi \CD{^\beta} \dch \notag\\
&+ b \dph \left( \frac{1}{2} \dg{_{\alpha\beta}} \CD{^\alpha} \chi \CD{^\beta} \chi
 - \frac{1}{4} \dg{^\alpha_\alpha} \CD{_\beta} \chi \CD{^\beta} \chi
 - \CD{_\alpha} \chi \CD{^\alpha} \dch \right)
 - \frac{1}{4} b^2 \dph^2 \CD{^\alpha} \chi \CD{_\alpha} \chi \notag \bigg)  \\
&+ \CD{_\mu} \bigg(e^{b\ph}\Big(
(\dg{_{\alpha \beta}} \dg{^{\mu \beta}} \chi \CD{^\alpha} \chi
-  \dg{^\mu_\alpha} \left( \dch \CD{^\alpha} \chi + \chi \CD{^\alpha} \dch \right) 
+ \dch \CD{^\mu}\dch)
- \frac{1}{2} \dg{^\beta_\beta} \dg{^\mu_\alpha} \chi \CD{^\alpha} \chi \notag\\
&- \frac{1}{4} \dg{_{\alpha\beta}} \dg{^{\alpha\beta}} \chi \CD{^\mu} \chi
+ \frac{1}{8} \dg{^\alpha_\alpha} \dg{^\beta_\beta} \chi \CD{^\mu} \chi
+ \frac{1}{2} \dg{^\alpha_\alpha} \left( \dch \CD{^\alpha} \chi + \chi \CD{^\alpha} \dch \right) \notag\\
&
+ b\dph(\chi \CD{^\mu} \dch - \dg{^\mu^\alpha} \chi \CD{_\alpha} \chi + \frac{1}{2} \dg{^\alpha_\alpha} \chi \CD{_\mu} \chi + \dch \CD{^\mu} \chi)
+\frac{1}{2} b^2 \dph^2 \chi \CD{^\mu}\chi\Big)\bigg)\bigg]
\end{align}
where the last three lines come from the boundary term. Equivalently, this can be obtained by inserting \eqref{eqn:perturbed-hodge-duality} in \eqref{eqn:perturbed-three-form-field-strength-axion-action}. 
%obtain the quadratic action associated with the scalar field formulation of the axion in \eqref{eqn:perturbed-scalar-field-axion-action} above, 
\footnote{Note that this requires keeping all contributions of the second-order perturbations of $\Ft{_{\nu\rho\sigma}}$ that contain products of $\dg{_{\mu\nu}}$ and the scalars $\dch, \dph$, which form a total derivative. Also, although contributions of $\delta^2 \Ft{_{\nu\rho\sigma}}$ are multiplied by background expressions that are zero, these must nevertheless be taken in account to show the equivalence of the perturbed action under the Hodge duality, since they lead to terms that include the product of two first-order perturbations.}

Expressing the three-form fluctuations as in \eqref{eqn:perturbed-three-form} and restricting to the scalar metric perturbations of \eqref{eqn:scalar-metric-perturbation}, the quadratic action \eqref{eqn:perturbed-three-form-field-strength-axion-action}, up to a total derivative on the induced metric, becomes 
\begin{align}\label{SE axion F}
\delta^2 S_E^{axion,F} = \int \dd{r} \dd[3]{x} \, \frac{\sqrt{\gamma}}{a^2} \bigg[& e^{-b \ph} \bigg( 
\frac{1}{2} \left( f^2 + \cd{_i} w \cd{^i} w \right) + Q \left( f \left( \dA + 3 \dpsi - \triangle \dE - b \dph  \right) + w \triangle \dB \right)\notag\\
&- \frac{1}{4} Q^2 \Big( 2 b \dph \left( \dA + 3 \dpsi - \triangle \dE \right) - b^2 \dph^2 + \dA^2 - 15 \dpsi^2  - 6 \dA \dpsi + 2 \dB \triangle \dB + 2 \dA \triangle \dE\notag\\
&+ 10 \dpsi \triangle \dE + \cd{_i} \dB \cd{^i} \dB - \triangle \dE \triangle \dE - 2 \cd{_i} \cd{_j} \dE \cd{^i} \cd{^j} \dE  \Big) \bigg)
 + 4 a^2 \dch \left( \Dot{f} - \triangle w \right) \bigg].
\end{align}
Likewise, the scalar form of the perturbed axion action \eqref{eqn:perturbed-scalar-field-axion-action}, excluding the final boundary term, equals
%restricted to axion fluctuations and select the scalar perturbations of the metric, the quadratic action in \eqref{eqn:perturbed-scalar-field-axion-action}, 
\begin{align}
\delta^2 S_E^{axion,\chi} = \int \dd{r} \dd[3]{x} \, a^2 \sqrt{\gamma}  e^{b\ph} \Big(&
\chid \left( \dchd \left( \dA + 3 \dpsi - \triangle \dE - b \dph \right) + \cd{_i} \dch \cd{^i} \dB \right)
- \frac{1}{2} \left( (\chid)^2 + \cd{_i} \chi \cd{^i} \chi \right) \notag\\
& - \frac{1}{4} (\chid)^2 \big( 2 b \dph \left( \dA + 3 \dpsi - \triangle \dE \right) + b^2 \dph^2 + 3 \dA^2 + 3 \dpsi^2 + 6 \dA \dpsi \notag\\
&- 2 \dpsi \triangle \dE  -2 \dA \triangle \dE + \cd{_i} \dB \cd{^i} \dB + \triangle \dE \triangle \dE - 2 \cd{_i} \cd{_j} \dE \cd{^i} \cd{^j} \dE \big)
\Big) .
\end{align}
Now, adding the boundary term contributions 
%that are a product of first order perturbations of the metric $\dg{_{\mu\nu}}$ and scalars $\dch, \dph$ 
and subtracting the contributions from the second-order perturbations $\Ft{_{\nu\rho\sigma}}$, we find that the difference is given by
% subtract the total derivative $\CD{_\mu} \left( \chi \, \eps{^{\mu\nu\rho\sigma}}\,  \delta^{(2)} \tensor{F}{_{\nu\rho\sigma}} \right) $
\begin{align}
\delta^2 S_E^\text{boundary} = \int \dd{r} \dd[3]{x} \, a^2 \sqrt{\gamma}  e^{b\ph} \Big(
(\dchd)^2 - \chid \dchd \left( \dA + 3 \dpsi -\triangle \dE \right) - \chid \cd{_i} \dch \cd{^i} \dB + \cd{_i} \dch \cd{^i} \dch
 + b \dph \chid \dchd
\Big).
\end{align}
Using the transformations derived from the Hodge duality rules \eqref{eqn:perturbed-hodge-duality}
\begin{align}\label{transformations axion}
\chid &\to \frac{e^{-b \ph} Q}{a^2},&
\dchd &\to \frac{e^{-b \ph}}{a^2} \left( f + Q \left( \dA + 3 \dpsi - \triangle \dE - b\dph \right) \right),
& \cd{_i} \dch &\to \frac{e^{-b \ph}}{a^2} \left( \cd{_i} w + Q \cd{_i} \dB \right),
\end{align}
we find that both formulations are equivalent, i.e.
\begin{equation}
\delta^2S_E^{axion,F}=\delta^2S_E^{axion,\chi}+\delta^2S_E^{boundary}.
\end{equation}

\subsection{Constraints and gauge invariance}\label{Gaugeinvformalism}
% 3.3
% Gauge invariant variables, boundary conditions and field equivalence
\subsubsection{Two-form formulation}
%When we combine the gravity sector with the field sector, we can simplify the resulting quadratic action by applying the equations of motion \eqref{eqn:equation-of-motion} and leaving aside several total derivatives with respect to $r$ or the induced metric. 
%The quadratic action still includes the term $\dch ( \Dot{f} - \triangle w  )$, which enforces the relation \eqref{dotf=deltaw}.
% Bruno's Thesis 5.15
% ACTION OF f and w, A, B, E, psi, ph, chi EIGHT scalar dof
\begin{comment}
\begin{align}
\delta^2 S_E = \int \dd{r} \dd[3]{x} \, \sqrt{\gamma} %\Bigg[ \frac{e^{-b\ph}}{a^2} \Big(&
\frac{1}{2} \left( f^2 - w \triangle w \right) 
+ Q \left( f \left(\dA + 3 \dpsi - \triangle \dE -b \dph \right) + \dB \triangle w \right)
+ \frac{1}{4} Q^2 b^2 \dph^2 \notag\\
&+ \frac{1}{2} Q^2 \left(
\left( \dA + 3 \dpsi - \triangle \dE \right)^2
- b \dph \left( \dA + 3 \dpsi - \triangle \dE \right)
- \dB \triangle \dB + w \triangle \dB
\right)
\Big)\notag\\
&+a^2 \Big( \frac{1}{2} \left( (\dphd)^2 - \dph \triangle \dph \right)
+ \phd \dph \left(3 \dpsid + \triangle \left( \dB - \dEd \right) \right) - \dA \phd \dphd \notag\\
&+ \dpsi \triangle \left(\dpsi - 2 \dA \right)
- 3 (\dpsid)^2 - 6 \Hh \dA \dpsid - 2 \left( \Hh \dA + \dpsid \right) \triangle \left( \dB - \dE'\right) \notag\\
&+ \kh \left( 3\dpsi^2 - 6 \dA \dpsi - 3 \dA^2 + \left(\dB - \dEd \right)\triangle \left(\dB - \dEd \right) \right)
\Big)
+ \dch \left( \Dot{f} - \triangle w\right)
\Bigg]
\end{align}
\end{comment}
The total perturbed action in the two-form formulation contains eight scalar degrees of freedom, given by the fields $\dA, \dB, \dpsi, \dE, \dph, \dch, f \text{ and } w$. However, \eqref{dotf=deltaw} relates $f$ and $w$, and Hodge duality in turn relates this to $\dch$, leaving us with six scalar degrees of freedom. The constraints and gauge freedom reduce this further to two physical degrees of freedom, as we now show.

%Now we will see what gauge invariance does. It cuts it down to two degrees of freedom eventually. To make this manifest we will search for constraint equations induced by non-dynamical variables. These non-dynamical variables are $A$ and $B$ and both enforce constraints that allows one to eliminate two variables in favour of the remaining two. How to do this will be dictated by gauge invariance. 

At this point it is both customary and convenient to switch to Fourier space, by expanding the scalar perturbation variables in spherical harmonics $Y_{nlm}$ that are eigenfunctions of the Laplacian $\triangle$ on $S^3$ with eigenvalues $-n^2+1$, with $n \in \mathbb{N}$. For simplicity, we drop the subscript $n$ on the fluctuation variables. 
%We suppress the indices that would distinguish the harmonics with the same eigenvalue. This is effectively equivalent to integrating out the $S^3$ coordinates, producing a factor of one when the harmonics $Y_n$ are properly normalised. 
%REMOVE: We will now treat each mode of $S^3$ separately, and for simplicity will not distinguish between $\triangle$ and its eigenvalue. 

Integrating out the Lagrange multiplier $\dch$ imposes the closure of $\Ft{}$ and allows us to replace $w$ with $\Dot{f}/\triangle$. Below we will compare the two-form fluctuation with the scalar one to find them equal. 

The combined quadratic action governing perturbation modes is then given by\footnote{Because of the $\Dot{f}/\triangle$ term, the action below seems to be badly-defined for $n=1$. As we will see, it becomes well-defined once all the constraints have been taken into account. Separate analysis of the $n=1$ mode starting from \eqref{SE axion F} with the constraint in \eqref{dotf=deltaw} just being given by $\Dot{f}=0$, yields the identical result. This separate analysis closely resembles the work in \cite{Jonas:2023qle}.}
\begin{align}
\delta^2 S_E = \int \dd{r} \dd[3]{x} \, \sqrt{\gamma} \bigg[ \frac{e^{-b\ph}}{a^2} \Big(&
\frac{1}{2} \left( f^2 - (\Dot{f})^2/\triangle \right) 
+ Q \left( f \left(\dA + 3 \dpsi - \triangle \dE -b \dph \right) + \dB \Dot{f} \right)
+ \frac{1}{4} Q^2 b^2 \dph^2 \notag\\
&+ \frac{1}{2} Q^2 \left(
\left( \dA + 3 \dpsi - \triangle \dE \right)^2
- b \dph \left( \dA + 3 \dpsi - \triangle \dE \right)
- \dB \triangle \dB + \dB \Dot{f}
\right)
\Big)
\notag\\
&+a^2 \Big( \frac{1}{2} \left( (\dphd)^2 - \dph \triangle \dph \right)
+ \phd \dph \left(3 \dpsid + \triangle \left( \dB - \dEd \right) \right) - \dA \phd \dphd \notag\\
&+ \dpsi \triangle \left(\dpsi - 2 \dA \right)
- 3 (\dpsid)^2 - 6 \Hh \dA \dpsid - 2 \left( \Hh \dA + \dpsid \right) \triangle \left( \dB - \dEd\right) \notag\\
&+ \kh \left( 3\dpsi^2 - 6 \dA \dpsi - 3 \dA^2 + \left(\dB - \dEd \right)\triangle \left(\dB - \dEd \right) \right)
\Big)
\bigg]
\end{align}
%Again, note that we found this action with the background field equations and an expansion into spherical harmonics.
%
To identify the constraints induced by the non-dynamical fields $\dA$ and $\dB$, we write the quadratic action in Hamiltonian form,
\begin{align}
S = \int \dd{r} \left[ p_i \Dot{q}_i - H(p,q) - C_i G^i(p,q)  \right],
\end{align}
where $C_i$'s are the Lagrange multipliers and $G^i(p,q)$ are the constraints they imply.

The conjugate momenta of the dynamical variables $\dpsi, \dE, f \text{ and } \dph$ are given by
\begin{align} % Bruno's Thesis 5.19
\Pf &= \frac{e^{-b\ph}}{\triangle a^2} \left( Q \triangle \dB - \Dot{f} \right),& \Ppsi &= a^2 \left(3 \phd \dphd - 6 \dA \Hh - 6 \dpsid +  2 \triangle\left( \dEd - \dB \right) \right), \notag\\
\Pdph &= a^2 \left( \dphd - \phd \dA \right),& \PE &= \triangle a^2 \left( - \phd \dphd + 2 \dA \Hh + 2 \dpsid + 2 \kh \left( \dEd - \dB \right) \right).
\end{align}
Substituting these in the action yields
\begin{align}
\delta^2 S_E =&\int \dd{r} \bigg[
\Ppsi \dpsid + \PE \dEd + \Pf \Dot{f} + \Pdph \dphd \notag \\
&+ \frac{1}{4 \left(\triangle + 3 \kh\right) a^2}
\left( \kh \Ppsi^2
- \frac{3}{\triangle} \PE^2
- 2 \Ppsi \PE \right)
+ \frac{a^2}{2} e^{b \ph} \triangle \Pf^2
+ \frac{1}{2a^2}\Pdph^2
- \frac{1}{2} \phd \Ppsi \dph 
+ a^2 \left(\triangle + 3 \kh \right) \psi^2\notag \\
&+ \frac{e^{-b\ph}}{2a^2} \left(f + Q (3 \dpsi - \triangle \dE - b\dph) \right)^2 - \frac{e^{-b\ph}}{4a^2} Q^2 b^2 \dph^2 
+ \frac{1}{2a^2} \left( \frac{3}{2} (\phd)^2 - \triangle \right) \dph^2
- \dB \left( \PE + Q \triangle \Pf \right)\notag\\
&- \dA \left( \Pdph \phd - \Hh \Ppsi + \frac{Q \, e^{-b \ph}}{a^2} \left( Q( \triangle \dE - 3 \dpsi + \frac{1}{2} b \dph ) - f \right)
+ a^2 \left( 3 \Hh \phd \dph + 2 ( \triangle + 3 \kh ) \dpsi \right) \right) \bigg]
\end{align}
which gives rise to the following two constraint equations, associated resp. with $\dA$ and $\dB$, 
\begin{align}
&\Pdph \phd - \Hh \Ppsi + \frac{Q \, e^{-b \ph}}{a^2} \left( Q( \triangle \dE - 3 \dpsi + \frac{1}{2} b \dph ) - f \right)
+ a^2 \left( 3 \Hh \phd \dph + 2 ( \triangle + 3 \kh ) \dpsi \right) = 0\\
&\PE + Q \triangle \Pf = 0.
\end{align}

Finally, to resolve the gauge redundancies we introduce the gauge-invariant variables
\begin{align}
R &= f - Q \triangle E,& \Phi &= \dph + \frac{\psi}{\Hh} \phd
\end{align}
Here $R$ is defined as in \cite{Loges:2022nuw} for the three-form fluctuations, and $\Phi$ is a gauge-invariant Mukhanov variable \cite{Garriga:1997wz,Gratton:2001gw}. Their conjugate momenta are
\begin{align}\label{actionRphi}
\PR &= \Pf + \frac{Q e^{-b \ph}}{a^2 \Hh}\psi,&
\PPh &= \Pdph - \left(\frac{b e^{-b \ph}}{2a^2 \Hh} + 3 a^2 \phd  \right) \dpsi.
\end{align}
The sought-after quadratic action in terms of $R$ and $\Phi$ is given by
\begin{align}\label{gauge inv three-form hamiltonian}
\delta^2 S_E =&\int \dd{r} \Bigg[\PR \Dot{R} + \PPh \Dot{\Phi}
+ \PR \left(
- \frac{e^{-b \ph} Q^2 \triangle}{2 a^4 \Hh (\triangle + 3 \kh)} R
+ \frac{Q \triangle}{4(\triangle+3\kh)} \left(\frac{b e^{-b\ph}Q^2}{a^4 \Hh} + 6 \phd \right) \Phi
\right)\notag\\
&+ \PPh \left(
\frac{Q \triangle \phd}{2a^2 \Hh \left(\triangle + 3 \kh \right)} \PR
- \frac{e^{-b\ph} Q \kh \phd}{2 a^4 \Hh^2 (\triangle + 3 \kh)} R
+ \frac{\phd\left(b e^{-b \ph} Q^2 \kh- 2 a^4 \Hh \triangle  \phd \right)}{4 a^4 \Hh^2 (\triangle + 3 \kh)} \Phi
\right)\notag\\
&+ \frac{\triangle \left(-3Q^2 + 2 e^{b \ph} a^4 (\triangle + 3 \kh) \right)}{4 a^2 (\triangle + 3 \kh)} \PR^2
+ \frac{e^{-2 b \ph} \left(-Q^2 \triangle + e^{b \ph} a^4 (\triangle + 3 \kh) \left(6 \kh + (\phd)^2\right)\right)}{12 a^6 \Hh^2 (\triangle+3\kh)} R^2\notag\\
&+ \frac{e^{-2b\ph} \left( b Q^3 (2 \triangle +3 \kh) - 2 e^{b\ph} Q a^4 \left( - 3 \Hh \triangle \phd + b (\triangle + 3 \kh) (6 \kh + (\phd)^2)  \right) \right)}{12 a^6 \Hh^2 (\triangle + 3 \kh) } R \Phi \notag\\
&+\frac{e^{-2 b \ph}}{48 a^6 \Hh^2 (\triangle + 3 \kh)}\Big( - b^2 Q^4 (2\triangle + 3 \kh) - e^{2b\ph} \triangle a^8 (6 \kh + (\phd)^2) (2 (\triangle+3\kh) + 3 (\phd)^2)\notag\\
&+e^{b \ph} Q^2 a^4 (2(\triangle + 3 \kh) (\triangle + 3 b^2 \kh) - 6 b \triangle \Hh \phd + (3 \triangle + b^2 \triangle+ 3 b^2 \kh)(\phd)^2) \Big) \Phi^2\notag\\
&+\frac{e^{-b\ph} \left(Q^2 (\triangle + 3 \kh) - e^{b \ph} a^4 (6 \kh (\triangle + 3 \kh)+ \triangle (\phd)^2)\right)}{12 a^6 \Hh^2 (\triangle + 3 \kh)} \PPh^2
\Bigg].
\end{align}
%This constitutes our sought-for quadratic action which depends only on two gauge-invariant variables and which has the constraints incorporated. 

\subsubsection{Scalar-field formulation}
A similar procedure yields the perturbed action in the scalar-field formulation. Combining the gravity with the field sector, and applying the equations of motion \eqref{eqn:equation-of-motion} we obtain, up to a total derivative in $r$ and a total derivative on the induced metric, the total perturbed action
\begin{align}
\delta^2 S_E =\int \dd{r} \dd[3]{x} a^2 \sqrt{\gamma} \Bigg[& e^{b \ph} \Big(
- \frac{1}{2} \left( \dchd^2 - \dch \triangle \dch \right) 
+ \chid \dchd \left( \dA - b \dph \right)
- \chid \dch \left( 3 \dpsid +\triangle ( \dB - \dEd \right)
+ \frac{1}{4} b \chid^2 \left( 2 \dA - b \dph \right) 
\Big) \notag\\
& +\frac{1}{2} \left( (\dphd)^2 - \dph \triangle \dph \right)
+ \phd \dph \left(3 \dpsid + \triangle \left( \dB - \dEd \right) \right) - \dA \phd \dphd \notag\\
&+ \dpsi \triangle \left(\dpsi - 2 \dA \right)
- 3 (\dpsid)^2 - 6 \Hh \dA \dpsid - 2 \left( \Hh \dA + \dpsid \right) \triangle \left( \dB - \dEd\right) \notag\\
&+ \kh \left( 3\dpsi^2 - 6 \dA \dpsi - 3 \dA^2 + \left(\dB - \dEd \right)\triangle \left(\dB - \dEd \right) \right)
\bigg]
\end{align}
The conjugate momenta of the dynamical variables $\psi, \dE, \chi \text{ and } \ph$ are 
\begin{align}
\Pdch &= e^{b \ph} a^2 \left( A \chid - b \dph \chid - \dchd \right) ,&
\Ppsi &= a^2 \left( 3 \phd \dphd - 3 e^{b \ph} \chid \dch - 6 \dA \Hh - 6 \dpsid + 2 \triangle\left( \dEd - \dB \right)  \right), \notag\\
\Pdph &= a^2 \left( \dphd - \phd \dA \right),&
\PE &= \triangle a^2 \left( - \phd \dphd - e^{b \ph} \chid \dchd + 2 \dA \Hh + 2 \dpsid + 2 \kh \left( \dEd - \dB \right) \right).
\end{align}
When substituted into the action, this yields the action in Hamiltonian form 
\begin{align}\label{hamiltonian scalar field}
\delta^2 S_E =&\int \dd{r} \bigg[\Ppsi \dpsid + \PE \dEd + \Pdch \dchd + \Pdph \dphd \notag\\
&+ \frac{1}{4 \left(\triangle + 3 \kh\right) a^2}
\left( \kh \Ppsi^2
- \frac{3}{\triangle} \PE^2
- 2 \Ppsi \PE \right)
+ \frac{1}{2 a^2} e^{- b \ph} \Pdch^2
+ \frac{1}{2 a^2}\Pdph^2
+ \Pdch b \dph \chid 
- \frac{1}{2} \Ppsi \left( \phd \dph - e^{b \ph} \chid \dch \right)\notag \\
&+ a^2 \left(\triangle + 3 \kh \right) \psi^2
+ \frac{1}{2} a^2 \left( \left( \triangle + \frac{3}{2} e^{b \ph} \chid^2 \right) \dch^2
+ \left(  \triangle - \frac{3}{2} \phd^2 - b^2 e^{b \ph} \chid^2 \right) \dph^2
- 3 e^{b \ph} \phd \dph \chid \dch \right)\notag\\
&- \dB \PE - \dA \left( \Pdph \phd + \Pdch \chid - \Hh \Ppsi +
+ a^2 \left( 3 \Hh \phd \dph - 3 \Hh e^{b \ph} \chid \dchd - \frac{1}{2} b e^{b \ph} \chid^2 + 2 ( \triangle + 3 \kh ) \dpsi \right) \right) \bigg]
\end{align}
and two constraints:
\begin{align}
& \Pdph \phd + \Pdch \chid - \Hh \Ppsi +
+ a^2 \left( 3 \Hh \phd \dph - 3 \Hh e^{b \ph} \chid \dchd - \frac{1}{2} b e^{b \ph} \chid^2 + 2 ( \triangle + 3 \kh ) \dpsi \right) = 0\\
& \PE = 0.
\end{align}
In terms of the gauge-invariant variables
\begin{align}
\Chi &= \dch + \frac{\psi}{\Hh} \chid, & \Phi &= \dph + \frac{\psi}{\Hh} \phd.
\intertext{with respective conjugate momenta }
\PCh &= \Pdch + 3 e^{b \ph} a^2 \chid \dpsi,&
\PPh &= \Pdph - \left(\frac{b e^{-b \ph}}{2a^2 \Hh} + 3 a^2 \phd  \right) \dpsi.
\end{align}
the quadratic action reads:
\begin{align}\label{scalar hamiltonian}
\delta^2 S_E =&\int \dd{r} \Bigg[\PCh \Dot{\Chi} + \PPh \Dot{\Phi}
+ \Chi \left(
  \frac{e^{b \ph} \chid^2 \triangle}{2 \Hh (\triangle + 3 \kh)} \PCh
+ \frac{e^{b \ph} \chid \triangle a^2}{4(\triangle+3\kh)} \left(\frac{b e^{b\ph} \chid^2}{\Hh} + 6 \phd \right) \Phi
\right)\notag\\
&+ \PPh \left(
\frac{e^{b \ph} \triangle \chid \phd}{2 \Hh \left(\triangle + 3 \kh \right)} \Chi
+ \frac{\kh \chid \phd}{2 a^2 \Hh^2 (\triangle + 3 \kh)} \PCh
+ \frac{\phd \left(b e^{b \ph} \chid^2 \kh - 2 \Hh \triangle  \phd \right)}{4 \Hh^2 (\triangle + 3 \kh)} \Phi
\right)\notag\\
&+ \frac{ e^{b \ph} a^2 \triangle \left( 2 (\triangle + 3 \kh) - 3 e^{b \ph} \chid^2 \right)}{4 (\triangle + 3 \kh)} \Chi^2
+ \frac{e^{- b \ph} \left( (\triangle + 3 \kh) \left(6 \kh + \phd^2 \right)-e^{b \ph} \triangle \chid^2 \right)}{12 a^2 \Hh^2 (\triangle+3\kh)} \PCh^2\notag\\
&-\frac{ \left( b e^{b\ph} \chid^2 (2 \triangle +3 \kh)  - 2 \chid \left( - 3 \Hh \triangle \phd + b (\triangle + 3 \kh) (6 \kh + (\phd)^2)  \right) \right)}{12 \Hh^2 (\triangle + 3 \kh) } \PCh \Phi \notag\\
&+\frac{a^2}{48 \Hh^2 (\triangle + 3 \kh)}\big( - b^2 e^{2 b \ph} \chid^4 (2\triangle + 3 \kh) - \triangle (6 \kh + (\phd)^2) (2 (\triangle+3\kh) + 3 (\phd)^2)\notag\\
&+ 2 e^{b\ph} \chid^2 (2(\triangle + 3 \kh) (\triangle + 3 b^2 \kh) - 6 b \triangle \Hh \phd + \phd^2 (3 \triangle + b^2 \triangle + 3 b^2 \kh)) \big) \Phi^2\notag\\
&-\frac{ \left( (\triangle + 3 \kh) (6 \kh -e^{b\ph} \chid^2) + \triangle (\phd)^2)\right)}{12 a^2 \Hh^2 (\triangle + 3 \kh)} \PPh^2
\Bigg].
\end{align}
By applying \eqref{transformations axion} and a symplectic transformation, 
\begin{align}
\Chi &\to \PR,& \PCh &\to - R,
\end{align}
including on the $\PCh \Chi$ boundary term, we recover exactly \eqref{gauge inv three-form hamiltonian}, as expected.

Remember also that eq. \eqref{gi boundary conditions} shows that our boundary conditions carry over straightforwardly to boundary conditions on the gauge-invariant variables. For the dilaton gauge-invariant variable $\Phi$ we have 
\begin{align}
\lim_{r \to \pm \infty} \Phi = \lim_{r \to \pm \infty}\left( \dph + \frac{\dpsi}{\Hh} \phd\right) = \dph,
\end{align}
Hence, Dirichlet boundary conditions on $\dph$ correspond to Dirichlet boundary conditions on $\Phi$.

Next, the condition that the axion charge remains unchanged at the boundary means that\footnote{Note that here we again refer to $\triangle$ as the Laplacian on the three-sphere, rather than the eigenvalue of the harmonic.}
\begin{align}
\delta Q_\text{axion} = \int_{S^3} \dd[3]{x} \sqrt{\gamma} \ f = \int_{S^3} \dd[3]{x} \sqrt{\gamma} ( R+ Q \triangle \dE ) =\int_{S^3} \dd[3]{x} \sqrt{\gamma} \ R
\end{align}
where the total derivative term integrates to zero. Therefore, the absence of axion-charge fluctuations translates to Dirichlet boundary conditions on $R$.

Finally, in the scalar-field formulation, the boundary term from the symplectic transformation implies a Neumann condition on $\Chi$, or equivalently, a Dirichlet condition on its conjugate momentum $\PCh$. Eq. \eqref{transformations axion} shows that this provides the correct boundary condition on the axion charge:
\begin{align}\label{charge fluctuations in terms of momenta}
\delta Q_\text{axion} = \int_{S^3} \dd[3]{x} \sqrt{\gamma} \left[ a^2 e^{b\ph} \dchd - Q\left( 3 \dpsi + \dA - \triangle \dE-b\dph \right) \right] = - \int_{S^3} \dd[3]{x} \sqrt{\gamma} \, \PCh,
\end{align}
Thus Dirichlet boundary conditions on axion-charge fluctuations correspond to Neuman conditions on $\Chi$.

\subsection{The choice of boundary conditions in \cite{Hertog:2018kbz}}

Before we analyse the spectrum of the quadratic action operator, we comment on the subtleties to do with the boundary conditions that have caused so much confusion in previous negative-mode studies.

In \cite{Hertog:2018kbz}, a stability analysis in the scalar-field formulation appeared to show that axion wormholes have multiple negative modes. By contrast, \cite{Loges:2022nuw} found that, based on an analysis in the three-form formalism, axion wormholes have no negative modes. This discrepancy can be traced to the erroneous choice of boundary conditions in \cite{Hertog:2018kbz} on the gauge-invariant variable that involves the axion. The reason this is a subtle issue is that this gauge-invariant variable is a combination of axion and metric perturbations, which are subject to respectively Neumann and Dirichlet boundary conditions.

To see this explicitly, consider the wormhole solution with the dilaton turned off, viz. $b=0$ and $\ph=0$. The axion gauge-invariant variable and its conjugate used in \cite{Hertog:2018kbz} were
\begin{equation}\label{gaugeinv43}
\begin{split}
    \dyg&=\frac{\Hh}{\chid}\Chi=\psi+\frac{\Hh}{\chid}\dch\quad, \\ 
    \Py&=\frac{\chid}{\Hh}\Pi_\Chi=\frac{\chid}{\Hh}\Pdch+\frac{3a^2\chid^2}{\Hh}\psi.
\end{split}
\end{equation}
In terms of these, the perturbed action reads
\begin{equation}\label{action43}
    \delta^2S_E= \int \diff r \left[\Py \Dot{\dyg}+ C_1 \Py ^2 + C_2\Py \dyg + C_3 \dyg^2\right]+\text{boundary terms},
\end{equation}
where
\begin{align}
        &C_1=\frac{\triangle \Hh^2 +3k^2(\triangle+3k)}{2 \chid^2a^2(\triangle+3k)},&
        &C_2=\frac{\kapd\triangle\dot{\chi}^2-4k(\triangle+3k)}{2(\triangle+3k)\mathcal{H}},&
        &C_3=\frac{\triangle a^2\dot{\chi}^2\left(2(\triangle+3k)-3\kapd\dot{\chi}^2\right)}{4(\triangle+3k)\mathcal{H}^2}.
\end{align} 
The fluctuation of axion charge in \eqref{charge fluctuations in terms of momenta} in terms of \eqref{gaugeinv43} is given by
\begin{equation}
    \delta Q_{\text{axion}} \ \propto \ \int_{S^3} \diff^3 x \sqrt{|\gamma|} \ a^2  \Py 
\end{equation}
Hence the boundary condition that the axion charge remains fixed requires that $\Py \to 0 $ faster than $e^{2|r|}$ in the asymptotic regions $|r| \to \infty$. It is safe to say indeed that one can obtain the results of \cite{Loges:2022nuw} by integrating out $\dyg$ in \eqref{action43}, defining the Sturm-Liouville problem of the form
\begin{equation}
    \hat{M}\Py= \lambda \Py,
\end{equation}
and calculating the eigenvalues of eigenfunctions that obey these boundary conditions. However, this was not what was done in \cite{Hertog:2018kbz}, where an analysis was performed based on an insufficiently precise adaptation of the results in \cite{Gratton:2001gw} to wormholes. 

Finally, we note that alternatively, one can integrate out the momenta $\Py$ from \eqref{action43}, which yields
\begin{equation}
     \delta^2S = \int \diff r\left[-\frac{1}{4C_1}\dot{\dyg}^2+\left(C_3-\frac{C_2^2}{4C_1}+\frac{\dif}{\dif r}\left(\frac{C_2}{4C_1}\right)\right)\dyg^2\right]-\left.\frac{C_2}{4C_1}\dyg^2\right|_\partial.
\end{equation}
Now, the crucial boundary term in this expression was omitted in \cite{Hertog:2018kbz}. Next a "symplectic flip" of the action was performed by defining the momentum conjugate of $\dyg$ as $\Pi=-\Dot{\dyg}/2C_1$, yielding
\begin{equation}
    \delta^2S_E= \int \diff r \left[K \Dot{\Pi}^2+V \Pi^2\right]
\end{equation}
for some functions $K$ and $V$. This expression led \cite{Hertog:2018kbz} to conclude that wormholes have infinitely many  fluctuation modes with Dirichlet boundary conditions on $\Pi$ that lower the action. However, given that
\begin{equation}
    \Pi=\Py+\frac{C_2}{2C_1}\dyg.
\end{equation}
it is now clear that 1) the axion charge does not remained unchanged under these fluctuations and 2) they obey Robin boundary conditions on $\dyg$, rather than Neumann conditions.

\section{Stability analysis}
To compute the spectrum of the quadratic action operator that governs the behaviour of linear perturbations around axion-saxion wormholes, we consider the action \eqref{gauge inv three-form hamiltonian} in the two-form formulation. 
Schematically this reads,
\begin{equation}\label{actionfunctions1}
\begin{split}
    \delta^2S_E=\int \diff r &\bigg[\Pi_\R\Dot{\R}+\Pi_\Phi \Dot{\Phi}+\A_n\Pi_\R^2+\B_n\Pi_\R\R+\C_n\R^2+\D_n \Pi_\Phi^2+\E_n\Pi_\Phi \Phi\\
    &+\F_n\Phi^2+\G_n\Pi_\Phi\Pi_\R+\HH_n\Pi_\R\Phi+\I_n \Pi_\Phi \R+\J_n \R \Phi\bigg],
\end{split}
\end{equation}
where we have denoted explicitly the dependence of the terms on the wavenumber $n$, because there are three rather distinct cases to consider:
\begin{itemize}
    \item  $n=1$ ($\triangle=0$). This is the homogeneous sector, for which \eqref{actionfunctions1} reduces to
\begin{equation}
    \delta^2S^{n=1}_E=\int \diff r \bigg[\Pi_\R\Dot{\R}+\Pi_\GS \Dot{\GS}+\C_1\R^2+\D_1 \Pi_\GS^2+\E_1\Pi_\GS \GS+\F_1\GS^2+\I_1 \Pi_\GS \R+\J_1 \R \GS\bigg].
\end{equation}
We see that $\Pi_\R$ is a Lagrange multiplier imposing $\Dot{\R}=0$. Hence homogeneous fluctuations are non-dynamical in the axion sector. Boundary conditions then imply that $\R=0$ \cite{Loges:2022nuw} and we are left with the dilaton sector, governed by
\begin{equation}\label{eq29}
    \delta^2S^{n=1}_E=\int \diff r \left[\Pi_\GS\Dot{\GS}+\D_1\Pi_\GS^2+\E_1\Pi_\GS\GS+\F_1\GS^2\right].
\end{equation}
To analyse the spectrum of this sector it proves convenient to bring \eqref{eq29} back into second-order form. Integrating out the momenta yields the following constraint
\begin{equation}
    \Dot{\GS}+2\D_1\Pi_\GS+\E_1\GS=0.
\end{equation}
which, when substituted, gives the second-order form of the action
\begin{equation}\label{hommodeaction}
    \delta^2 S^{n=1}_E=\int \diff r \left[-\frac{1}{4\D_1}\Dot{\GS}^2-\frac{\E_1}{2\D_1}\Dot{\GS}\GS+\left(\F_1-\frac{\E_1^2}{4\D_1}\right)\GS^2\right],
\end{equation}
where
\begin{equation}\label{hommodefunctions}
\begin{split}
    &\D_1=\frac{e^{-b\varphi}\kapd Q^2-6a^4k}{12a^6\mathcal{H}^2}, \quad \E_1=\frac{be^{-b\varphi}\kapd Q^2 \Dot{\varphi}}{12a^4\mathcal{H}^2}, \\
    &\F_1=\frac{b^2e^{-2b\varphi}Q^2(-\kapd Q^2+2e^{b\varphi}a^4(6k+\kapd\Dot{\varphi}^2))}{48a^6\mathcal{H}}.
\end{split}
\end{equation}
Hence in contrast with axion wormholes without a dilaton field turned on, the dilaton renders the homogeneous mode dynamical. We discuss this sector below in Section \ref{homogeneous mode chapter}. 

\item  $n=2$ ($\triangle=-3k$):  All the terms in \eqref{actionfunctions1} are divergent. This mode is non-dynamical, since it requires infinite action to excite \cite{Gratton:1999ya}. That said, \cite{Loges:2022nuw} argued that for axion wormholes, $\psi$ and $E$ form a new dynamical field $\psi+E$. However, it turns out that the action of this mode is a total-derivative term which moreover vanishes because of the boundary conditions, leading back to the conclusion that this mode is non-dynamical.
\end{itemize}

For all other values of $n$, fluctuations are dynamical in both the axion and dilaton sectors. Again, we impose Dirichlet boundary conditions on $\R$ and $\GS$, thus we have to integrate out their momenta in order to bring the action back into second-order form. From the action in \eqref{actionfunctions1}, we can easily see that integrating out momenta will impose the following linear constraints
\begin{equation}\label{constraintsmomenta}
    \begin{pmatrix}
        2 \A_n & \G_n\\
        \G_n & 2 \D_n
    \end{pmatrix}\begin{pmatrix}
        \Pi_\R \\
        \Pi_\GS 
    \end{pmatrix}=-\begin{pmatrix}
        \Dot{\R}+\B_n\R+\HH_n \GS\\
        \Dot{\GS}+\E_n\GS+\I_n \R
    \end{pmatrix}.
\end{equation}
This system of equations is invertible if $4\A_n \D_n-\G^2_n \neq 0$ which, written out explicitly, says
\begin{equation}
    \frac{\triangle}{12 a^8\mathcal{H}^2}\bigg ( \big (2 e^{b\varphi}a^4-\frac{3\kapd Q^2}{\triangle+3k}\big)\big (e^{-b\varphi}\kapd Q^2-\frac{a^4(6k(\triangle+3k)+\kapd\triangle \Dot{\varphi}^2}{\triangle+3k}\big)-\frac{3\kapf Q^2\triangle a^4\Dot{\varphi}^2}{(\triangle+3k)^2}\bigg)\neq0.
\end{equation}
Substituting the explicit background solutions for the dilaton and scale factor gives
\begin{equation}
%There is a kappa to be inserted here%
\begin{split}
&\frac{\triangle  Q^2 \kapf \lvert c \rvert}{72 a^8\mathcal{H}^2(\triangle+3k)^2}\left[(\triangle+3k)\cos^2\bigg(\sqrt{\frac{3}{2\kapd}}b\arctan (\tanh (r))\bigg)\cosh^2(2r)-9\right]\\
 &\left[6k+\triangle+\triangle \cos \bigg (\sqrt{6}b\arctan (\tanh (r))\bigg)\sec \bigg (\sqrt{\frac{3}{2\kapd}}b \arctan (\tanh (r))\bigg)^2-2k (\triangle+3k)\right]\neq 0.
\end{split}
\end{equation}
Note that the first bracket is never zero, since $\triangle+3k<0$, whereas the second bracket is zero at the wormhole throat $r=0$ when $k=1$, which is our case of interest. 
\renewcommand{\A}{\mathcal{A}_n}
\renewcommand{\B}{\mathcal{B}_n}
\renewcommand{\C}{\mathcal{C}_n}
\renewcommand{\D}{\mathcal{D}_n}
\renewcommand{\E}{\mathcal{E}_n}
\renewcommand{\F}{\mathcal{F}_n}
\renewcommand{\G}{\mathcal{G}_n}
\renewcommand{\HH}{\mathcal{I}_n}
\renewcommand{\I}{\mathcal{J}_n}
\renewcommand{\J}{\mathcal{K}_n}

Upon integrating out the momenta, we obtain the following action 
    \begin{equation}\label{2action}
    \delta^2S_E = \int \diff r\left[\A\dot{\R}^2+\B \R\dot{\R}+\C \R^2+\D\dot{\GS}^2+\E\dot{\GS}\GS+\F\GS^2+\G\R\GS + \HH \dot{\R}\GS+\I\R\dot{\GS}+\J\dot{\R}\dot{\GS}\right],
\end{equation}
where $\A,\dots,\J$ are the functions obtained after inverting \eqref{constraintsmomenta}. We give their explicit expressions in Appendix \ref{functionsofthesturmliouvilleproblem}. 

In fact, since $\A\sim 1/a^2$ as $r\rightarrow \pm \infty$, it proves convenient to work with the rescaled perturbation variables
\begin{equation}\label{rescaling}
    \R(r)=\sqrt{\cosh{2r}}\ \PP(r).
\end{equation}
The modified set of functions in the action \eqref{2action} after this rescaling are given in Eq \eqref{functions after rescaling} in Appendix \ref{functionsofthesturmliouvilleproblem}. 
Asymptotically,
\begin{equation}
    \R(r)\xrightarrow{\lvert r\rvert \rightarrow\infty} e^{\lvert r\rvert} \ \PP(r).
\end{equation}
Hence the rescaled perturbation $\PP(r)$ must decay faster than $e^{-\lvert r\rvert }$ asymptotically to meet our boundary condition. 
Finally, writing \eqref{2action} as 
\begin{equation}
    \delta^2S_E=\int \diff r \begin{pmatrix}
    \PP & \GS
    \end{pmatrix}\hat{\M}\begin{pmatrix}
    \PP \\
    \GS
    \end{pmatrix}+\mathrm{boundary}\ \mathrm{terms},
\end{equation}
gives us the matrix Sturm-Liouville problem
\begin{equation}
    \hat{\M}\begin{pmatrix}
        \PP \\
        \GS
    \end{pmatrix}=\lambda
    \begin{pmatrix}
        \PP \\
        \GS
    \end{pmatrix},
\end{equation}
where $\hat{\M}$ is a self-adjoint matrix differential operator whose matrix entries are
\begin{equation}\label{sturmliouvillematrixelements}
\begin{split}
    \M_{11}&=-\frac{\diff}{\diff r}\left (\A\frac{\diff}{\diff r}\right)+\C-\frac{1}{2}\dot{\B}, \\ \M_{12}&=-\frac{1}{2}\J\frac{\diff^2}{\diff r^2}+\frac{1}{2}(\I-\HH-\dot{\J})\frac{\diff}{\diff  r}+\frac{1}{2}\left(\G-\dot{\HH}\right),\\
    \M_{21}&=-\frac{1}{2}\J\frac{\diff^2}{\diff r^2}+\frac{1}{2}(\HH-\I-\dot{\J})\frac{\diff}{\diff r}+\frac{1}{2}\left(\G-\dot{\I}\right), \\
    \M_{22}&=-\frac{\diff}{\diff r}\left (\D\frac{\diff}{\diff r}\right)+\F-\frac{1}{2}\dot{\E}.
\end{split}
\end{equation}
The boundary terms in the action are explicitly given by
\begin{equation}
\bigg \{\A \PP \Dot{\PP}+\frac{\B}{2}\PP^2+\D\GS \Dot{\GS}+\frac{\E}{2}\GS^2+\frac{\J}{2}(\PP\Dot{\GS}+\Dot{\PP}\GS)+\frac{1}{2}(\HH+\I)\PP \GS\bigg \} \bigg \rvert_\partial.
\end{equation}
The only non-trivial boundary term is $\D \GS \Dot{\GS}$, since asymptotically $\D\sim a^2$. This means that $\GS$ must decay faster than $e^{-\lvert r\rvert}$ at the boundary. Thus we obtain a separate Sturm-Liouville problem for each $S^3$ mode $n>2$, which we now analyse. 
\subsection{Homogeneous sector}\label{homogeneous mode chapter}
The action \eqref{hommodeaction} of the homogeneous dilaton sector takes the form 
\begin{equation}\label{homomodeaction1}
    \delta^2 S_E=\int \diff r\left[ \alpha \Dot{\GS}^2+\beta \Dot{\GS}\GS+\gamma \GS^2\right],
\end{equation}
where 
\begin{equation}\label{function hom mode2}
\begin{split}
    \alpha&=\frac{\sqrt{| c| } \sinh ^2(2 r) \cosh (2 r)}{2 \sqrt{6} \left(\cosh ^2(2 r)-\sec ^2\left(\sqrt{\frac{3}{2}} b \arctan(\tanh (r))\right)\right)},\\
    \beta&=\frac{b \sqrt{| c| } \tan \left(\sqrt{\frac{3}{2}} b \arctan(\tanh (r))\right) \sec ^2\left(\sqrt{\frac{3}{2}} b \arctan (\tanh (r))\right)}{2 \sec ^2\left(\sqrt{\frac{3}{2}} b \arctan(\tanh (r))\right)-2 \cosh ^2(2 r)},\\
    \gamma&=\frac{\sqrt{\frac{3}{2}} b^2 \sqrt{| c| } \text{sech}(2 r) \sec ^2\left(\sqrt{\frac{3}{2}} b \arctan(\tanh (r))\right) \left(\sec ^2\left(\sqrt{\frac{3}{2}} b \arctan(\tanh (r))\right)-2 \cosh ^2(2 r)\right)}{4 \left(\cosh ^2(2 r)-\sec ^2\left(\sqrt{\frac{3}{2}} b \arctan(\tanh (r))\right)\right)}.
\end{split}
\end{equation}
\begin{figure}
    \centering
    \subfloat[$\alpha$]{\includegraphics[width=0.3\textwidth]{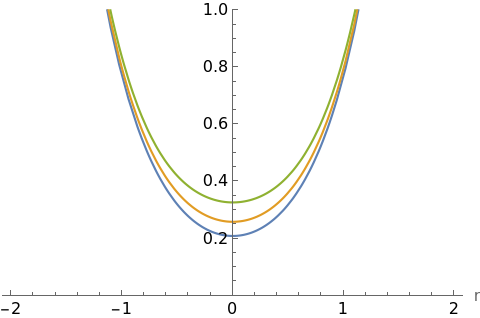}}\hspace{20pt}
    \subfloat[$\gamma-\frac{\dot{\beta}}{2}$]{\includegraphics[width=0.3\textwidth]{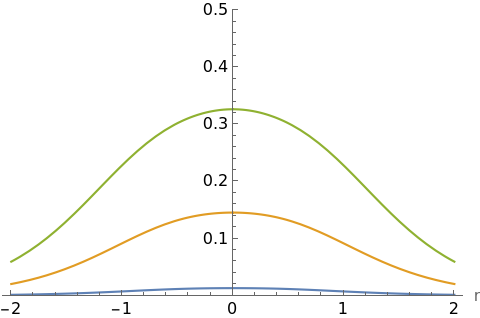}}\hspace{20pt}
    \subfloat{\includegraphics[width=0.075\textwidth]{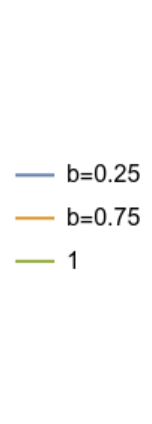}}
    \caption{Functions in the homogeneous mode action \eqref{afterintegrationhom}, for different values of $b$ (with $c=-1$).}
\label{functionshommode}
\end{figure}
Not surprisingly, the axion charge $Q$ doesn't enter in this action. Note also that the coefficients $\alpha,\beta$ and $\gamma$ depend on $c$ only via a multiplicative factor $\sqrt{\lvert c \rvert}$. Hence the dependence on the size of the wormhole is rather trivial. The only non-trivial physical parameter is the dilaton coupling $b$.

Now, since $\beta\sim e^{-4\lvert r\rvert}$ asymptotically, we can integrate the $\beta \Dot{\Phi}\Phi$ term in \eqref{homomodeaction1} by parts to obtain
\begin{equation}\label{afterintegrationhom}
    \delta^2 S_E=\int \diff r \left[\alpha \Dot{\Phi}^2+\left(\gamma-\frac{\dot{\beta}}{2}\right)\Phi^2\right]+\frac{\beta}{2}\Phi^2\bigg\rvert_\partial.
\end{equation}
The boundary term vanishes since $\GS$ obeys Dirichlet boundary conditions. Further, Figure \ref{functionshommode} shows that the coefficient functions \eqref{afterintegrationhom} are everywhere positive. Hence homogeneous fluctuations are suppressed. There is no negative mode in this sector, given the boundary conditions of interest.  

A numerical analysis of the Sturm-Liouville problem defined from \eqref{homomodeaction1} confirms this. Note that even though the functions $\beta$ and $\gamma$ are singular at the wormhole throat, the combination entering in the Sturm-Liouville problem is well-behaved, as shown in Figure \ref{functionshommode}. The singular behaviour does mean, however, that only odd eigenfunctions contribute to the homogeneous sector, since even eigenfunctions give rise to infinite action. 

Note that we did not have to perform any additional Hawking-Perry rotation in our analysis to suppress the homogeneous fluctuations. This shows that we do not encounter the conformal factor problem.
   \subsection{Spectral analysis of the inhomogeneous modes}\label{inhomogeneous modes chapter}
The inhomogeneous $n>2$ modes are governed by the Sturm-Liouville problem defined in \eqref{sturmliouvillematrixelements}. The coupling terms mean that we must resort to a numerical analysis.
As before, even though many of the individual functions in the quadratic action \eqref{2action} are singular at the wormhole throat, the combinations of the functions that enter in the Sturm-Liouville problem are well-behaved. They are shown in Appendix \ref{functionsofthesturmliouvilleproblem} in Figure \ref{slfunctions}. 

Once again, therefore, this means that only odd eigenfunctions contribute to the physically relevant spectrum of fluctuations. 
\begin{figure}[t!]
\centering
 \subfloat[$n=3$]{\includegraphics[width=0.45\textwidth]{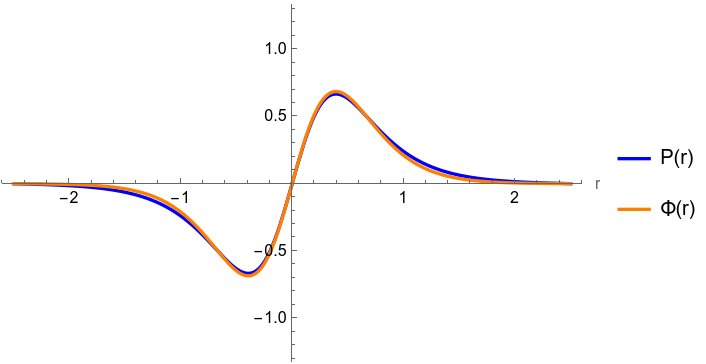}}\hspace{20pt}  \subfloat[$n=4$]{\includegraphics[width=0.45\textwidth]{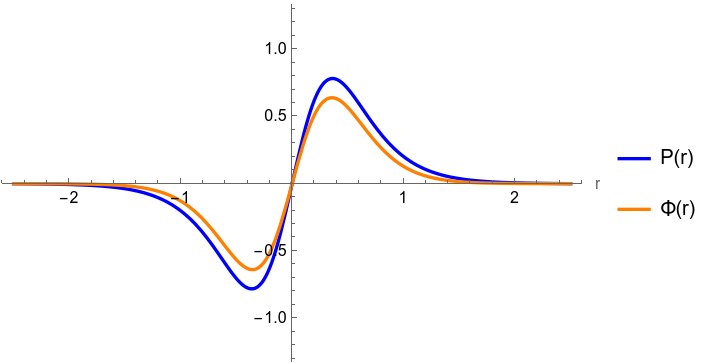}}\\
\caption{The two lowest odd eigenfunctions of the Sturm-Liouville problem ($n=3,4$) in an axion-dilaton wormhole background, for $Q=0.5$, $b=1$ and $c=-1$.}
\label{dileigenfunctions}
\end{figure}

To see that even eigenfunctions have infinite action, consider their contribution to the action from the near-throat region. Since even eigenfunctions are approximately constant near $r=0$, the action integral across a narrow interval $ 
\Delta r = 2\epsilon$ across the throat is approximately 
\begin{equation}\label{eveneigenf}
    \delta^2S_E \approx \int_{-\epsilon}^\epsilon \diff r \left[\C \PP_e^2+\F \GS_e^2+\G \PP_e \GS_e\right]\ ,
\end{equation}
where $\PP_e,\GS_e$ are constants and $C,F,G$ behave as $1/r^2$ near $r=0$.
In fact, since
\begin{equation}
    \lim_{r\to 0} \bigg \lvert \frac{\G}{\sqrt{\C\F}} \bigg \lvert \to 2.
\end{equation}
it turns out that \eqref{eveneigenf} is approximately 
\begin{equation}
    \delta^2S_E \approx \int_{-\epsilon}^\epsilon \diff r \left(\sqrt{\C}\PP_e\pm\sqrt{\F}\GS_e\right)^2 > 0.
\end{equation}
Hence all even eigenfunctions are infinitely suppressed in the path integral.

We now turn to odd eigenfunctions. The spectrum of this Sturm-Liouville operator is found using the shooting method. Since odd eigenfunctions behave linearly very near the origin, we consider the small-$r$ ansatz 
\begin{equation}
    \PP(r)\sim \PP_0 r, \quad \GS \sim \GS_0 r, 
\end{equation}
for some constants $\PP_0$ and $\GS_0$. Next we perform a shooting procedure out to the asymptotic region, where we implement the correct boundary conditions. This shooting procedure is slightly more involved than usual and is explained in detail in Appendix \ref{numerics}. Here we present the main results. The two lowest (normalized) odd eigenfunctions\footnote{For each $n$ we have a Sturm-Liouville problem on its own and hence, its own spectrum. It should not come as a surprise then that for all $n>2$, the lowest odd eigenfunction contains only one node. Same result was found in \cite{Loges:2022nuw} for axion wormholes.} for $Q=0.5$ are shown in Figure \ref{dileigenfunctions}. The eigenvalues for the five lowest modes are given in Table \ref{eigenvalues:axdil}, for different values of $Q$. 

We see the eigenvalues are all positive, so we can conclude that axion-dilaton wormholes are perturbatively stable.
 \begin{table}[b!]
\begin{center}
\centering
\begin{tabular}{||c | c c | c c| c c ||} 
 \hline
  & \multicolumn{2}{|c|}{$Q=0.5$} & \multicolumn{2}{|c|}{$Q=1$}  & \multicolumn{2}{|c|}{$Q=2.5$} \\ 
  \hline\hline
$S^3$ Mode  &  $\lambda$& $\PP_0/\GS_0$  &  $\lambda$ &  $\PP_0/\GS_0$ &   $\lambda$ &  $\PP_0/\GS_0$   \\ [0.5ex] 
 \hline
 $n=3$ &  1.98270 & 0.98598  &  0.736703 & 2.46824 &  0.130886 & 6.76735  \\
 $n=4$  & 3.18903 & 1.21699   & 1.04043 & 3.27718 & 0.176521 & 9.03553\\
 $n=5$ &  4.06330 & 1.59486 &   1.19782 & 4.31269  &  0.198302 & 11.7600 \\
 $n=6$ & 4.64923 & 2.11251 &  1.28512 & 5.55172  & 0.209958 & 14.9410 \\ 
 $n=7$ & 5.01677 & 2.76031  & 1.33514 & 6.98657 & 0.216504 & 18.5827 \\[1ex]  
 \hline 
 \hline
\end{tabular}
\caption{Eigenvalues and the values of the ratio $\PP_0/\GS_0$ for which the asymptotic boundary conditions hold, for five different $S^3$ modes and three values of $Q$, with $b=1$ and $c=-1$.}
\label{eigenvalues:axdil}
\end{center}
\end{table}
\begin{figure}[t!]
\centering
 \subfloat[$\lambda$]{\includegraphics[width=0.35\textwidth]{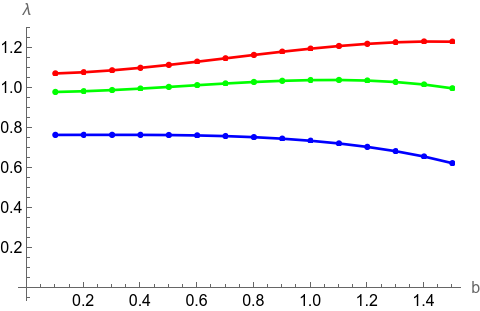}}\hspace{40pt}
 \subfloat[$\PP_0/\GS_0$]{\includegraphics[width=0.35\textwidth]{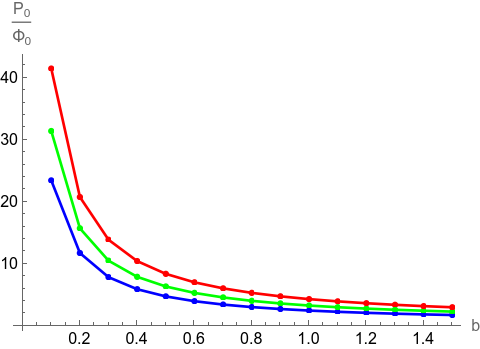}}
\caption{The eigenvalues $\lambda$ and the ratio $\PP_0/\GS_0$) as a function of the dilaton coupling $b$, for $n=3$ (blue), $n=4$ (green) and $n=5$ (red). Other parameters are fixed in this Figure, $Q=1$ and $c=-1$.}
\label{figuresvscoupling}
\end{figure}

Figure \ref{figuresvscoupling} illustrates how the eigenvalues $\lambda$ and the ratio $\PP_0/\Phi_0$ depend on the dilaton coupling $b$. (The dependence of $\lambda$ on $Q$ and $c$ is given in Appendix \ref{dependence appendix}).
For $b \rightarrow 0$, $\PP_0/\GS_0$ diverges, indicating that axion fluctuations dominate in this limit. In this regime, we can ignore the dilaton fluctuations and the spectrum should agree with the spectrum in axion wormhole backgrounds without dilaton. We confirmed that this is indeed the case. The spectrum of the purely axionic Sturm-Liouville problem agrees with that found in this section in the limit of $b\rightarrow 0$. For consistency, we also verified that our results agree with those in \cite{Loges:2022nuw}.

\section{Conclusions}
We have demonstrated that spherically symmetric Euclidean wormholes sourced by an axion and dilaton field are perturbatively stable. This generalizes and corrects earlier studies. Further, we made explicit how the stability analyses carried out in both Hodge-dual frames are equivalent. Given the subtleties to do with the sign of the kinetic term of the axion scalar upon Wick rotation and Hodge duality, this resolves a certain degree of confusion in the existing literature and elucidates the final result that these solutions are perturbatively stable. 

An obvious extension of this work includes an analysis of the stability of wormholes in general sigma models. We believe this is fairly straightforward, since more general wormhole backgrounds of this kind are described by geodesics on the target space. This means one could make use of a  local coordinate frame adapted to geodesic curves, the so-called free-falling coordinates. Another extension concerns a stability analysis of wormholes in the presence of a non-zero cosmological constant. For a positive cosmological constant, \cite{Aguilar-Gutierrez:2023ril} showed that axionic wormholes without a dilaton field are perturbatively stable. Given the similarities between wormhole solutions in flat space and in AdS, we similarly expect that our stability results remain qualitatively unchanged in the presence of a negative cosmological constant. Finally, it would be interesting to consider the effect of dilaton masses, which would be relevant in a phenomenological setting where supersymmetry is broken \cite{Andriolo:2022rxc, Jonas:2023ipa}\footnote{In AdS, it is possible for the dilaton to have mass even when SUSY is retained \cite{Loges:2023ypl}.}. 
%Whether these extensions leave room for instabilities is unclear and we believe that our simple analysis at least has taken away the intuition that negative modes could be centered around the wormhole.  
In this context \cite{Jonas:2023qle} recently demonstrated that several classes of wormhole solutions with dilaton potentials are perturbatively stable, at least with respect to homogeneous perturbations. It would be natural to extend their formalism to non-homogeneous fluctuations.

Evidently the absence of negative modes has consequences for the interpretation of these wormholes as saddle points in the gravitational path integral. In order to resolve the factorization paradox in AdS/CFT it would seem that all wormhole contributions should exactly cancel out. This is known as the ``baby universe hypothesis'' \cite{Marolf:2020xie, McNamara:2020uza}. Further, for Euclidean axion wormholes one may expect that this cancellation should happen within each superselection sector separately, labelled by the discrete axion fluxes $Q$. The perturbative stability of axion wormholes, even in the presence of a dilaton, means that it remains very much unclear how this is realized.

\section*{Acknowledgements}
%\begin{acknowledgments}
%This work is supported by the KU Leuven grant C16/16/005 - Horizons in hoge-energie-fysica.

We thank  Caroline Jonas, Gary Shiu, Gregory J. Loges, and Sergio E. Aguilar for various very valuable discussions. S.M. also thanks Manuel Kr\"amer for guidance during his first master thesis when he initially started on this topic. The research of T.H., R.T., and T.V.R. is in part supported by the Odysseus grant GCD-D5133-G0H9318N of FWO-Vlaanderen and KU Leuven C1 grant ZKD1118 C16/16/005. T.H. and S.M. acknowledge support from the inter-university project (IBOF/21/084).

%\end{acknowledgments}

\appendix

\section{Matrix elements of the two-field Sturm-Liouville problem}\label{functionsofthesturmliouvilleproblem}
\begin{figure}[p] % Use [p] to put the figure on a separate page
    \centering
    \subfloat[$\A$]{\includegraphics[width=0.3\textwidth]{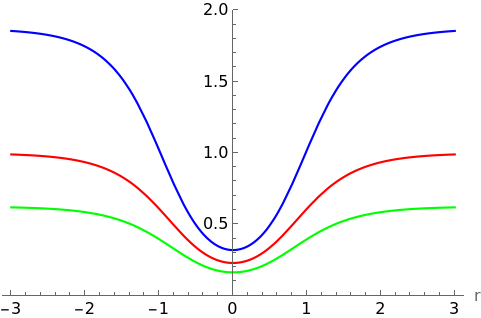}}\hspace{20pt}
    \subfloat[$\C-\frac{1}{2}\Dot{\B}$]{\includegraphics[width=0.3\textwidth]{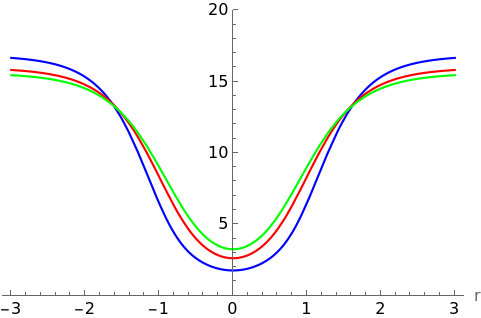}}\hspace{20pt}\\
    
    \subfloat[$\D$]{\includegraphics[width=0.3\textwidth]{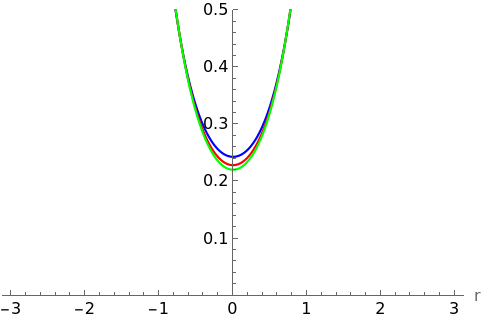}}\hspace{20pt}
    \subfloat[$\F-\frac{1}{2}\Dot{\E}$]{\includegraphics[width=0.3\textwidth]{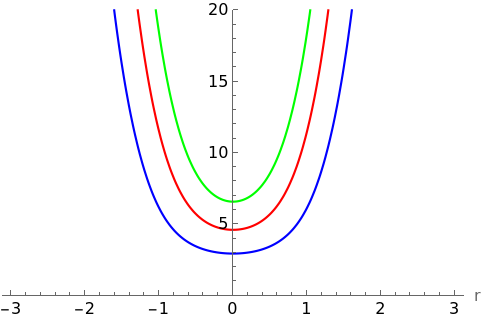}}\hspace{20pt}\\
    
    \subfloat[$\frac{1}{2}(\HH-\I-\dot{\J})$]{\includegraphics[width=0.3\textwidth]{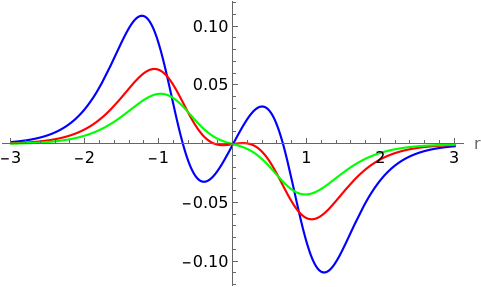}}\hspace{20pt}
    \subfloat[$\frac{1}{2}(\I-\HH-\dot{\J})$]{\includegraphics[width=0.3\textwidth]{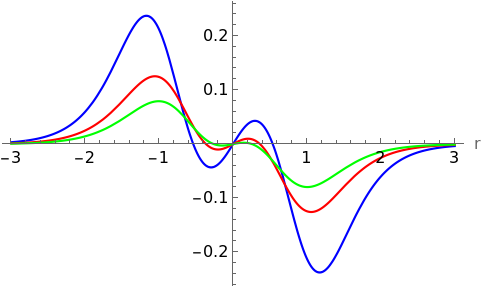}} \\
    
    \subfloat[$-\frac{1}{2}\left(\G-\dot{\HH}\right)$]{\includegraphics[width=0.3\textwidth]{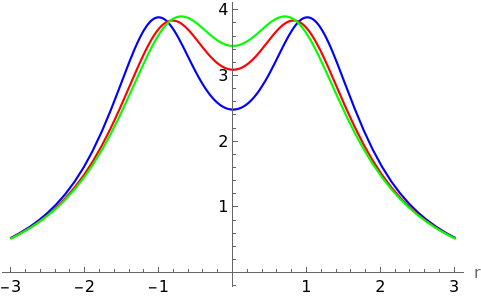}}\hspace{20pt}
    \subfloat[$-\frac{1}{2}\left(\G-\dot{\I}\right)$]{\includegraphics[width=0.3\textwidth]{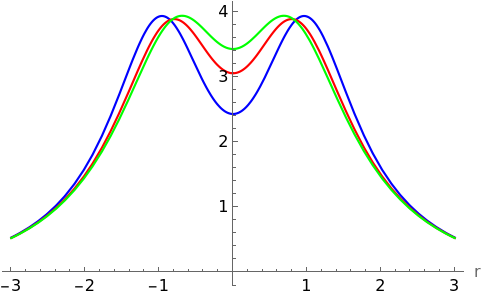}}\hspace{20pt}
    \subfloat[-$\J$]{\includegraphics[width=0.3\textwidth]{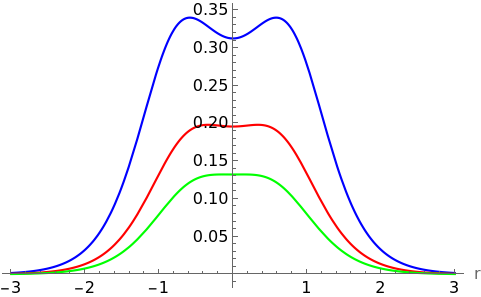}}
    
    \caption{Functions involved in the matrix elements of the Sturm-Liouville problem in \eqref{sturmliouvillematrixelements}, after rescaling \eqref{rescaling} for $n=3$ (blue), $n=4$ (red) and $n=5$ (green), where $Q=0.5$, $b=1$ and $c=-1$.}
    \label{slfunctions}
\end{figure}

In this appendix we present the exact expressions for the functions in the Sturm-Liouville problem in \eqref{sturmliouvillematrixelements} 
\begin{equation}
    \begin{split}
        \A&=\frac{\kapd(\triangle+3k)Q^2a^2-e^{b\varphi}a^6(6k(\triangle+3k)+\kapd\triangle\dot{\varphi}^2)}{\triangle(3\kapf Q^4-2\kapd e^{b\varphi}(\triangle+12k)Q^2a^4+2e^{2b\varphi}a^8(6k(\triangle+3k)+\kapd\triangle\dot{\varphi}^2))}\\
        \B&=\frac{6\kapd Q^2a^2\mathcal{H}}{3\kapf Q^4-2e^{b\varphi}\kapd(\triangle+12k)Q^2a^4+2e^{2b\varphi}a^8(6k(\triangle+3k)+\kapd\triangle\dot{\varphi}^2)}\\
        \C&=\frac{e^{-b\varphi}(-\kapd Q^2+e^{b\varphi}a^4(6k+\kapd\dot{\varphi}^2))(-\kapd(\triangle+9k)Q^2+e^{b\varphi}a^4(6k(\triangle+3k)+\kapd\triangle\dot{\varphi}^2)}{6a^2\mathcal{H}^2(3\kapf Q^4-2e^{b\varphi}\kapd(\triangle+12k)Q^2a^4+2e^{2b\varphi}a^8(6k(\triangle+3k)+\kapd\triangle\dot{\varphi}^2))}\\
        \D&=\frac{e^{-b\varphi}(-3\kapd Q^2+2e^{b\varphi}(\triangle+3k)a^4)(\kapd Q^2-e^{b\varphi}a^4(6k+\kapd\dot{\varphi}^2))^2}{12a^2\mathcal{H}^2(3\kapf Q^4-2e^{b\varphi}\kapd(\triangle+12k)Q^2a^4+2e^{2b\varphi}a^8(6k(\triangle+3k)+\kapd\triangle\dot{\varphi}^2))}\\
        \E&=\frac{e^{-b\varphi}\kapd\Dot{\varphi}(b\kapf Q^6(\triangle+3k)-4e^{3b\varphi}\triangle a^{12}\mathcal{H}(\triangle+3k)\Dot{\varphi}(6k+\kapd\Dot{\varphi}^2)}{4a^2\mathcal{H}^2(\triangle+3k)(3\kapf Q^4-2e^{b\varphi}\kapd Q^2a^4(\triangle+12k)+2e^{2b\varphi}a^8(6k(\triangle+3k)+\kapd\triangle\Dot{\varphi}^2))}\\
        & -\frac{e^{b\varphi}\kapd Q^4a^4(6\kapd\triangle\mathcal{H}\Dot{\varphi}+b(\triangle+3k)(8k+\kapd\Dot{\varphi}))}{4a^2\mathcal{H}^2(\triangle+3k)(3\kapf Q^4-2e^{b\varphi}\kapd Q^2a^4(\triangle+12k)+2e^{2b\varphi}a^8(6k(\triangle+3k)+\kapd \triangle\Dot{\varphi}^2))}\\
        &+\frac{2e^{2b\varphi}Q^2a^8(-18\kapd\triangle\mathcal{H}^3\Dot{\varphi}+bk(\triangle+3k)(6k+\kapd\Dot{\varphi}^2)+\kapd\triangle\mathcal{H}\Dot{\varphi}(2\triangle+24k+3\kapd\Dot{\varphi}^2))}{4a^2\mathcal{H}^2(\triangle+3k)(3\kapf Q^4-2e^{b\varphi}\kapd Q^2a^4(\triangle+12k)+2e^{2b\varphi}a^8(6k(\triangle+3k)+\kapd\triangle\Dot{\varphi}^2))}\\
        \G&=e^{-2b\varphi}\bigg (\frac{(3b\kaps Q^7(\triangle+3k)-4e^{3b\varphi}Qa^{12}(\triangle+3k)(6k+\kapd\Dot{\varphi}^2(6bk(\triangle+3k)-3\kapd\triangle\mathcal{H}\Dot{\varphi}+b\kapd\triangle\Dot{\varphi}^2)}{12a^6\mathcal{H}^2(\triangle+3k)(3\kapf Q^4-2e^{b\varphi}\kapd Q^2a^4(\triangle+12k)+2e^{2b\varphi}a^8(6k(\triangle+3k)+\kapd\triangle\Dot{\varphi}^2))}\\
        &-\frac{e^{b\varphi}\kapf Q^5a^4(-18\kapd\triangle\mathcal{H}\Dot{\varphi}+b(\triangle+3k)(4\triangle+60k+3\kapd\Dot{\varphi}^2))}{12a^6\mathcal{H}^2(\triangle+3k)(3\kapf Q^4-2e^{b\varphi}\kapd Q^2a^4(\triangle+12k)+2e^{2b\varphi}a^8(6k(\triangle+3k)+\kapd\triangle\Dot{\varphi}^2))}\\    &+\frac{2e^{2b\varphi}\kapd Q^3a^8(54\kapd\triangle\mathcal{H}^3\Dot{\varphi}-3\kapd\triangle\mathcal{H}\Dot{\varphi}(2\triangle+24k+3\kapd\Dot{\varphi})}{12a^6\mathcal{H}^2(\triangle+3k)(3\kapf Q^4-2e^{b\varphi}\kapd Q^2a^4(\triangle+12k)+2e^{2b\varphi}a^8(6k(\triangle+3k)+\kapd\triangle\Dot{\varphi}^2))}\\
        &+\frac{b(\triangle+3k)(6k(4\triangle+27k)+\kapd(4\triangle+21k)\Dot{\varphi^2}))}{12a^6\mathcal{H}^2(\triangle+3k)(3\kapf Q^4-2e^{b\varphi}\kapd Q^2a^4(\triangle+12k)+2e^{2b\varphi}a^8(6k(\triangle+3k)+\kapd\triangle\Dot{\varphi}^2))}\bigg)\\
        \HH&=\frac{3\kapd Qa^2 \mathcal{H}(b\kapd Q^4-36e^{2b\varphi}a^8\mathcal{H}k\Dot{\varphi}-e^{b\varphi}Q^2a^4(6bk-6\kapd\mathcal{H}\Dot{\varphi}+b\kapd\Dot{\varphi}^2))}{(\kapd Q^2-e^{b\varphi}a^4(6k+\kapd\Dot{\varphi}^2))(3\kapd Q^4-2e^{b\varphi}\kapd Q^2a^4(\triangle+12k)+2e^{2b\varphi}a^8(6k(\triangle+3k)+\kapd\triangle\Dot{\varphi}^2))}\\
        \I&=\frac{e^{-b\varphi}\kapd Q(\kappa_4^2Q^2-2e^{b\varphi}a^4k)\dot{\varphi}(\kapd Q^2-e^{b\varphi}a^4(6k+\kapd \dot{\varphi}))}{2a^2\mathcal{H}^2(3\kapf Q^4-2e^{b\varphi}\kapd(\triangle+12k)Q^2a^4+2e^{2b\varphi}a^8(6k(\triangle+3k)+\kapd\triangle\dot{\varphi}^2))}\\
        \J&=\frac{6e^{b\varphi}\kapd Qa^6\mathcal{H}\dot{\varphi}}{3\kapd Q^4-2e^{b\varphi}\kapd Q^2a^4(\triangle+12k)+2e^{2b\varphi}a^8(6k(\triangle+12k)+\kapd\triangle\Dot{\varphi}^2)}.\\
    \end{split}
\end{equation}
The function $\F$ is rather complicated and we write it separately
\begin{equation}
    \begin{split}
        \F&=\frac{1}{\mathcal{N}}e^{-2b\varphi}\bigg(-3b^2\kaps Q^8(2\triangle^2+9k\triangle+9k^2)+e^{b\varphi}\kapd Q^6a^4(\triangle+3k)(-18b^2\triangle\mathcal{H}^2+2(6\kapd\triangle(\triangle+3k)\\
        &+b^2(2\triangle^2+45\triangle k+90k^2))-36b\kapd\triangle\mathcal{H}\Dot{\varphi}+3\kapd(6\kapd\triangle+b^2(4\triangle+3k))\Dot{\varphi}^2\\
        &-4e^{4b\varphi}\triangle a^{16}(\triangle+3k)\big(72k^2(\triangle+3k)^2+24\kapd k(\triangle^2+9\triangle k+18k^2)\Dot{\varphi}^2\\
        &+2\kapd(\triangle^2-9\triangle\mathcal{H}^2+21\triangle k+ 27 k^2)\Dot{\varphi}^4+3\kaps\triangle \Dot{\varphi}^6\big)\\
        &-2e^{2b\varphi}\kapd Q^4a^8\big(2(\triangle+3k)^2(2\kapd\triangle(\triangle+21k)+3b^2k(4\triangle+27k))+108b\kapd\triangle \mathcal{H}^3(\triangle+3k)\Dot{\varphi}\\
        &-18\kapd(b^2+3\kapd)\triangle\mathcal{H}^2(\triangle+3k)\Dot{\varphi}^2+\kapd(\triangle+3k)(18\kapd\triangle(\triangle+8k)+b^2(4\triangle^2+51\triangle k+63k^2))\Dot{\varphi}^2\\
        &+3\kapd\triangle(b^2(\triangle+3k)+3\kapd(2\triangle+3k))\Dot{\varphi}^2-6b\kapd\triangle\mathcal{H}(\triangle+3k)\Dot{\varphi}(2\triangle+24k+3\kapd\Dot{\varphi}^2))\\
        &+2e^{3b\varphi}Q^2a^{12}(24k(\triangle+3k)^2(3b^2k(\triangle+3k)+\kapd\triangle(2\triangle+15k))\\
        &+4\kapd(\triangle+3k)(3b^2k(2\triangle^2+9\triangle k+9k^2)+3\kapd\triangle(2\triangle^2+45\triangle k +171 k^2))\Dot{\varphi}^2+2\kapf\triangle(b^2(\triangle+3k)\\
        &+9\kapd(\triangle^2+10\triangle k+12k^2))\Dot{\varphi}^2+9\kape\triangle^2\Dot{\varphi}^6-12b\kapd\triangle\mathcal{H}(\triangle+3k)^2\Dot{\varphi}(6k+\kapd\Dot{\varphi}^2)\\
        &-18\kapf\triangle\mathcal{H}^2\Dot{\varphi}^2(2\triangle^2+24\triangle k +54 k^2+3\kapd\triangle\Dot{\varphi}^2)))\bigg),
    \end{split}
\end{equation}
where 
\begin{equation}
    \mathcal{N}=48a^6\mathcal{H}^2(\triangle+3k)^2(3\kapf Q^4-2e^{b\varphi}\kapd Q^2a^4(\triangle+12k)+2e^{2b\varphi}a^8(6k(\triangle+3k)+\kapd\triangle\Dot{\varphi}^2)).
\end{equation}
After the rescaling of the gauge-invariant perturbations \eqref{rescaling} the former functions change to
\begin{equation}\label{functions after rescaling}
\begin{split}
   &\A \ \rightarrow \ \A \cosh{2r}, \quad \B \ \rightarrow \ \B \cosh{2r}+2\A \sinh{2r},\\
   &\C \ \rightarrow\ \C \cosh{2r}+\B \sinh{2r}+\A\sinh{2r}\tanh{2r}\\,
   &\G \ \rightarrow \ \sqrt{\cosh{2r}}\ \G +\frac{\sinh{2r}}{\sqrt{\cosh{2r}}}\ \HH,\\
   &\HH\ \rightarrow \ \sqrt{\cosh{2r}}\HH, \quad \I \ \rightarrow \ \sqrt{\cosh{2r}}\ \I +\frac{\sinh{2r}}{\sqrt{\cosh{2r}}}\ \J, \quad \J \ \rightarrow \sqrt{\cosh{2r}}\ \J.
\end{split}
\end{equation}
$\D, \E, \F$ remain unchanged under this rescaling of perturbations. They are given in Figure \ref{slfunctions}.

\newpage
\section{Numerical method for the odd eigenfunctions}\label{numerics}
In this appendix we explain the numerical method used to calculate the spectrum in Section \ref{inhomogeneous modes chapter} for odd eigenfunctions. Odd eigenfunctions exhibit a linear behaviour near the origin, so we consider the small-$r$ ansatz
\begin{equation}
    \PP(r)\sim \PP_0 r, \quad \GS \sim \GS_0 r, 
\end{equation}
for some constants $\PP_0$ and $\GS_0$ and then perform the shooting method to infinity. In the asymptotic regions $r\rightarrow\pm\infty$ there will be a combination of decaying and diverging solutions, we want to make sure that the diverging ones vanish consistent with Dirichlet boundary conditions. In order to find asymptotic solutions, we first solve the decoupled differential equations
\begin{equation}\label{solsatinfinity}
    -\frac{\diff}{\diff r}\left (\A \frac{\diff}{\diff r}\right)\PP+\left(\C-\frac{1}{2}\Dot{\B}\right)\PP=\lambda \PP, \quad -\frac{\diff }{\diff r}\left(\D\frac{\diff}{\diff r}\right)\GS+\left(\F-\frac{1}{2}\Dot{\E}\right)\GS=0.
\end{equation}
There is no term proportional to the eigenvalue in the equation for $\GS$ since $\D$ and $\left(\F-\frac{1}{2}\Dot{\E}\right)$ behave as $e^{2\lvert r\rvert}$ asymptotically and they are dominant over that term. The off-diagonal terms arising from the coupling act as sources for the particular solutions when solutions to \eqref{solsatinfinity} are substituted in.
If we define
\begin{equation}
    \alpha_1=\sqrt{n^2-a_0\lambda}, \quad \alpha_2=\sqrt{1+b_0}-1.
\end{equation}
where 
\begin{equation}
    a_0=\lim_{\lvert r\rvert\rightarrow \infty}\frac{1}{\A}=(n^2-1)\sqrt{\frac{2\lvert c \rvert }{3}}e^{b\varphi}\bigg \rvert_{\lvert r\rvert\rightarrow\infty}\quad b_0=\lim_{\lvert r\rvert\rightarrow \infty}\frac{1}{\D}\left(\F-\frac{1}{2}\Dot{\E}\right)=n^2-1,
\end{equation}
we can simply write the diverging solutions as 
\begin{equation}\label{atinf}
\begin{split}
    &\PP(r)\xrightarrow{\lvert r \rvert \rightarrow \infty} \beta_1\exp(\alpha_1\lvert r\rvert)+ \beta_2\PP^{\mathrm{div}}_{\mathrm{part}}(\lvert r\rvert), \\
    &\GS(r)\xrightarrow{\lvert r \rvert \rightarrow \infty}\beta_2\exp(\alpha_2\lvert r \rvert)+\beta_1\GS^{\mathrm{div}}_{\mathrm{part}}(\lvert r\rvert),
\end{split}
\end{equation}
where $\beta$'s are integration constants. The term multiplied with an integration constant $\beta_2$ appears also in the asymptotic expression for $\PP$ because, since $\beta_2 \exp (\alpha_2 \lvert r \rvert)$ is the solution of the decoupled equation in \eqref{solsatinfinity}, it will act as a source for the particular solution in $\PP$ thus leading to the term proportional to $\beta_2$. Similar reasoning holds for the appearance of $\beta_1$ in the asymptotic expression for $\GS$.

These particular solutions are given by
\begin{equation}\label{decaysol1}
   \PP^{\mathrm{div}}_{\mathrm{part}}(r)=  c_1 e^{(\alpha_2-1)r}+c_2 e^{(\alpha_2-3) r}, \quad    \GS^{\mathrm{div}}_{\mathrm{part}}(r)=d_1 e^{(\alpha_1-3) r}+d_2 e^{(\alpha_1-5) r},
\end{equation}
where $c_i$ and $d_i$ are coefficients which are determined numerically. The decaying solutions in the asymptotic regions are
\begin{equation}\label{decayingsoln}
\begin{split}
    &\PP(r)\xrightarrow{\lvert r \rvert \rightarrow \infty} \gamma_1\exp\left(-\alpha_1\lvert r\rvert\right)+\gamma_2\PP^{\mathrm{dec}}_{\mathrm{part}}(\lvert r\rvert), \\
    &\GS(r) \xrightarrow{\lvert r \rvert \rightarrow \infty} \gamma_2 \exp\left(-\left(\alpha_2+2\right)\lvert r \rvert \right)+\gamma_1\PP^{\mathrm{dec}}_{\mathrm{part}}(\lvert r\rvert),
\end{split}
\end{equation}
where $\gamma$'s are integration constants. The argument for their appearance in both of these asymptotic expressions is identical to the one presented for $\beta$'s above. As one can see from \eqref{decayingsoln}, there are new particular solutions induced by decaying solutions and they are given by\footnote{The particular solutions in \eqref{atinf} and \eqref{decayingsoln} are not the only ones we can include, e.g. we can do another iteration so that $\GS_{\mathrm{part}}^\lambda$ can now induce another solution for $\PP$ and vice versa. However, asymptotic solutions included in \eqref{atinf} and \eqref{decayingsoln} will always contain the most dominant behaviour in the asymptotic regions, thus it is not necessary to include any further solutions that may be induced.}
\begin{equation}\label{partdec}
   \PP^{\mathrm{dec}}_{\mathrm{part}}(r)=  c_3 e^{-(\alpha_2+3)r}+c_4 e^{-(\alpha_2+5) r}, \quad    \GS^{\mathrm{dec}}_{\mathrm{part}}(r)=d_3 e^{-(\alpha_1+3) r}+d_4 e^{-(\alpha_1+5) r},
\end{equation}
where $c_{3,4}$ and $d_{3,4}$ are again determined numerically.

To perform an appropriate shooting method in two-dimensional Sturm-Liouville problems, we need two parameters which are varied in the shooting. We want integration constants in \eqref{atinf} to be equal to $\beta_1=\beta_2=0$, thereby eliminating the divergent solutions. In the shooting code itself, those integration constants are defined as functions of $r$ by inverting \eqref{atinf} such that they approach a constant value asymptotically. There is a precise combination of constants coming from the behaviour near $r=0$, which we will denote as ($\PP_0,\GS_0$), and the eigenvalue $\lambda$ which makes them vanish. Moreover, for each value of the parameters ($Q,n\dots$), there will be a $\lambda$ for which a certain ratio of $\PP_0/\GS_0$ will eliminate the diverging solutions and impose vanishing boundary conditions on the fluctuations. The extra shooting parameter, along with the eigenvalue $\lambda$, represents the ratio of slopes originating from the linear behaviour of the two gauge-invariant fluctuations near the wormhole throat. This should not come as a surprise. Since the fluctuations are coupled, one could initially expect the importance of a mutual hierarchy between the fluctuations, which makes them source each other out and vanish at infinity.
\begin{figure}[t!]
\centering
 \subfloat[$n=3$]{\includegraphics[width=0.35\textwidth]{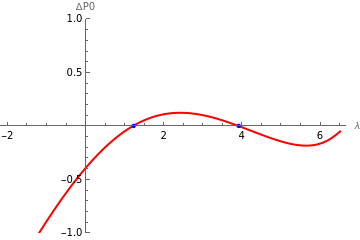}}\hspace{30pt}
 \subfloat[$n=4$]{\includegraphics[width=0.35\textwidth]{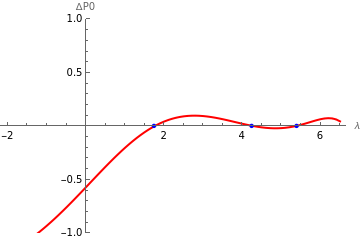}}
\caption{Function $\Delta \PP_0(\lambda)$ as defined in the text, zeroes of this function correspond to the spectrum of the respective inhomogeneous mode. However, one should check that the zeroes do in fact represent the eigenvalues and that they are not an artefact of numerical errors. Sketched for $Q=0.75$, $b=1$ and $c=-1$.}
\label{deltap}
\end{figure}
Since it is the ratio of $\PP_0$ and $\Phi_0$ that matters in the shooting method, we have a freedom in setting one of the constants to some value, we choose $\GS_0=1$, and then find the appropriate values ($\lambda,\PP_0$) which eliminate the diverging solutions. This just means that $\GS_0$ has been absorbed in the normalization. The eigenfunctions need to be orthonormal and actual values of $\PP_0$ and $\GS_0$ are then determined by normalization. 

We perform the shooting in the following way: for a certain $\lambda$ we identify
\begin{equation}
\begin{split}
    &\PP_0^1: \quad \text{value of } \PP_0 \text{ that makes } \beta_1=0, \\
    &\PP_0^2: \quad \text{value of } \PP_0 \text{ that makes } \beta_2=0, 
\end{split}
\end{equation}
with the FindRoot function in Mathematica. This defines $\Delta \PP_0= \PP_0^1-\PP_0^2$, which depends on the eigenvalue, i.e. $\Delta \PP_0(\lambda)$, and it is sketched in Figures \ref{deltap} for the lowest inhomogeneous modes. We then identify the values $\lambda_a$ and $\lambda_b$ for which $\Delta \PP_0(\lambda)$ has different signs. In the interval $[\lambda_a,\lambda_b]$ we perform a linear interpolation to find the eigenvalue
\begin{equation}
    \lambda_c=\lambda_b-\Delta \PP_0(\lambda_b)\frac{\lambda_a-\lambda_b}{\Delta \PP_0(\lambda_a)-\Delta \PP_0(\lambda_b)},
\end{equation}
over a couple of iterations to make $\Delta \PP_0\approx 0$.\footnote{Alternatively, one can introduce another FindRoot function which then determines the value of $\lambda$ for which $\Delta \PP_0(\lambda)\approx 0$.} This determines the eigenvalue and the value of $\PP_0$. As discussed previously, there will be an upper bound on $\lambda$ due to the fact that $\PP(r)$ must decay faster than $e^{-\lvert r\rvert }$. This would translate to the bound $ \alpha_1<1$ or
\begin{equation}
    \lambda < \sqrt{\frac{3}{2\lvert c \rvert}}e^{-b\varphi}\bigg \rvert_{\lvert r \rvert \rightarrow \infty}=\sqrt{\frac{3 \lvert c \rvert}{2}} \frac{1}{Q^2\cos^2\left(\frac{\pi}{4}\sqrt{\frac{3}{2}}b \right)}.
\end{equation}
With this bound satisfied, fluctuations also decay consistently as demanded by the non-trivial boundary term $\D$.

These results can also be confirmed by usual shooting from infinity where the initial conditions are taken as follows. For a certain value of $\lambda$, the decaying solutions in \eqref{decayingsoln} will give the initial conditions in terms of integration constants $\gamma_1$ and $\gamma_2$. Again, it is the ratio that matters since one of the constants can be absorbed in the normalization, thus we can set one of them equal to one, e.g. $\gamma_2=1$. Interpolation is then done as before to determine the value of $\gamma_1$ which eliminates the diverging solutions in the other asymptotic region. The spectrum and the values of $\PP_0/\GS_0$ agree with Table \ref{eigenvalues:axdil}.

\newpage
\section{Dependence of shooting parameters on $Q$ and $c$}\label{dependence appendix}
\begin{figure}[h!]
\centering
 \subfloat[$\lambda$]{\includegraphics[width=0.35\textwidth]{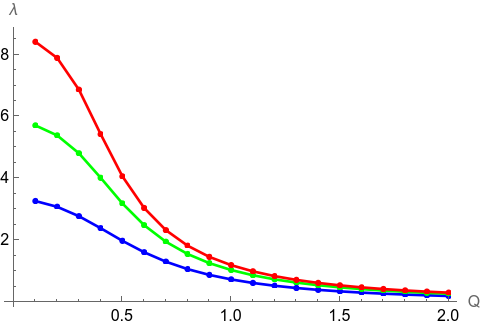}}\hspace{40pt}
 \subfloat[$\PP_0/\GS_0$]{\includegraphics[width=0.35\textwidth]{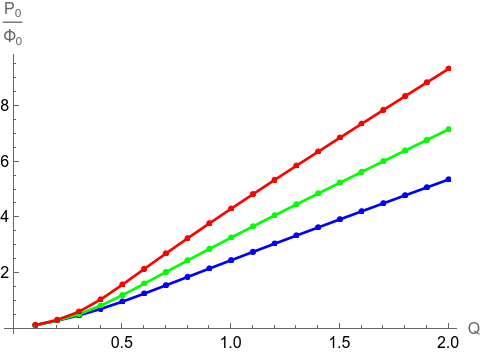}}
\caption{Dependence of the parameters determined with shooting method ($\lambda$ and $\PP_0/\GS_0$) on axion charge $Q$ for $n=3$ (blue), $n=4$ (green) and $n=5$ (red). Other parameters are fixed and set to be $b=1$ and $c=-1$.}
\label{figuresvscharge}

 \subfloat[$\lambda$]{\includegraphics[width=0.35\textwidth]{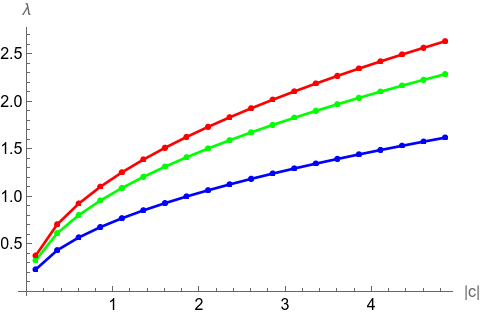}}\hspace{40pt}
 \subfloat[$\PP_0/\GS_0$]{\includegraphics[width=0.35\textwidth]{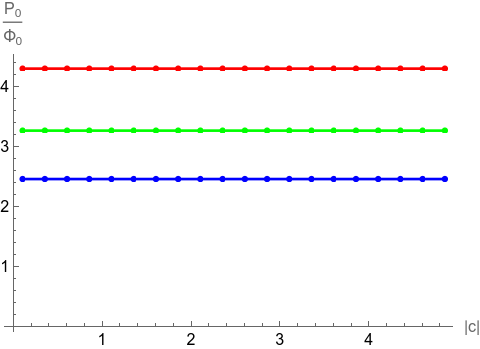}}
\caption{Dependence of the parameters determined with shooting method ($\lambda$ and $\PP_0/\GS_0$) on wormhole size, which is determined by $\lvert c \rvert$, for $n=3$ (blue), $n=4$ (green) and $n=5$ (red). Other parameters are fixed and set to be $b=1$ and $Q=1$.}
\label{figuresvssize}
\end{figure}

In this appendix we show how $\lambda$ and $\PP_0/\GS_0$ from Section \ref{inhomogeneous modes chapter} depend on the parameters $Q$ and $c$. This is outlined in Figures \ref{figuresvscharge} and \ref{figuresvssize}. Combining the results from Table \ref{eigenvalues:axdil} and Figure \ref{figuresvscharge} done for the lowest inhomogeneous modes, we can see that the value of $\PP_0/\GS_0$ increases with the increase of axion charge, which is sensible since $\PP$ is related to charge fluctuations. Figure \ref{figuresvssize} shows the dependence on $\lvert c \rvert$. Since this parameter is proportional to the size of the wormhole, we can see that the eigenvalues are larger for macroscopic wormholes, while $\PP_0/\Phi_0$ has the same value for both microscopic and macroscopic wormholes. 

\bibliographystyle{toine}

\bibliography{refs}

\providecommand{\href}[2]{#2}\begingroup\raggedright\begin{thebibliography}{10}

\bibitem{Hertog:2018kbz}
T.~Hertog, B.~Truijen, and T.~Van~Riet, ``{Euclidean axion wormholes have
  multiple negative modes},''
  \href{http://dx.doi.org/10.1103/PhysRevLett.123.081302}{{\em Phys. Rev.
  Lett.} {\bfseries 123} no.~8, (2019) 081302},
  \href{http://arxiv.org/abs/1811.12690}{{\ttfamily arXiv:1811.12690
  [hep-th]}}.

\bibitem{Giddings:1987cg}
S.~B. Giddings and A.~Strominger, ``{Axion Induced Topology Change in Quantum
  Gravity and String Theory},''
  \href{http://dx.doi.org/10.1016/0550-3213(88)90446-4}{{\em Nucl. Phys. B}
  {\bfseries 306} (1988) 890--907}.

\bibitem{Giddings:1988cx}
S.~B. Giddings and A.~Strominger, ``{Loss of Incoherence and Determination of
  Coupling Constants in Quantum Gravity},''
  \href{http://dx.doi.org/10.1016/0550-3213(88)90109-5}{{\em Nucl. Phys. B}
  {\bfseries 307} (1988) 854--866}.

\bibitem{Giddings:1989bq}
S.~B. Giddings and A.~Strominger, ``{STRING WORMHOLES},''
  \href{http://dx.doi.org/10.1016/0370-2693(89)91651-1}{{\em Phys. Lett. B}
  {\bfseries 230} (1989) 46--51}.

\bibitem{Coleman:1988cy}
S.~R. Coleman, ``{Black Holes as Red Herrings: Topological Fluctuations and the
  Loss of Quantum Coherence},''
  \href{http://dx.doi.org/10.1016/0550-3213(88)90110-1}{{\em Nucl. Phys. B}
  {\bfseries 307} (1988) 867--882}.

\bibitem{Lavrelashvili:1987jg}
G.~V. Lavrelashvili, V.~A. Rubakov, and P.~G. Tinyakov, ``{Disruption of
  Quantum Coherence upon a Change in Spatial Topology in Quantum Gravity},''
  {\em JETP Lett.} {\bfseries 46} (1987) 167--169.

\bibitem{Hebecker:2018ofv}
A.~Hebecker, T.~Mikhail, and P.~Soler, ``{Euclidean wormholes, baby universes,
  and their impact on particle physics and cosmology},''
  \href{http://dx.doi.org/10.3389/fspas.2018.00035}{{\em Front. Astron. Space
  Sci.} {\bfseries 5} (2018) 35},
  \href{http://arxiv.org/abs/1807.00824}{{\ttfamily arXiv:1807.00824
  [hep-th]}}.

\bibitem{Martucci:2024trp}
L.~Martucci, N.~Risso, A.~Valenti, and L.~Vecchi, ``{Wormholes in the axiverse,
  and the species scale},'' \href{http://arxiv.org/abs/2404.14489}{{\ttfamily
  arXiv:2404.14489 [hep-th]}}.

\bibitem{Andriolo:2022rxc}
S.~Andriolo, G.~Shiu, P.~Soler, and T.~Van~Riet, ``{Axion wormholes with
  massive dilaton},'' \href{http://dx.doi.org/10.1088/1361-6382/ac8fdc}{{\em
  Class. Quant. Grav.} {\bfseries 39} no.~21, (2022) 215014},
  \href{http://arxiv.org/abs/2205.01119}{{\ttfamily arXiv:2205.01119
  [hep-th]}}.

\bibitem{Jonas:2023ipa}
C.~Jonas, G.~Lavrelashvili, and J.-L. Lehners, ``{Zoo of axionic wormholes},''
  \href{http://dx.doi.org/10.1103/PhysRevD.108.066012}{{\em Phys. Rev. D}
  {\bfseries 108} no.~6, (2023) 066012},
  \href{http://arxiv.org/abs/2306.11129}{{\ttfamily arXiv:2306.11129
  [hep-th]}}.

\bibitem{Jonas:2023qle}
C.~Jonas, G.~Lavrelashvili, and J.-L. Lehners, ``{Stability of axion-dilaton
  wormholes},'' \href{http://dx.doi.org/10.1103/PhysRevD.109.086022}{{\em Phys.
  Rev. D} {\bfseries 109} no.~8, (2024) 086022},
  \href{http://arxiv.org/abs/2312.08971}{{\ttfamily arXiv:2312.08971
  [hep-th]}}.

\bibitem{Andriolo:2020lul}
S.~Andriolo, T.-C. Huang, T.~Noumi, H.~Ooguri, and G.~Shiu, ``{Duality and
  axionic weak gravity},''
  \href{http://dx.doi.org/10.1103/PhysRevD.102.046008}{{\em Phys. Rev. D}
  {\bfseries 102} no.~4, (2020) 046008},
  \href{http://arxiv.org/abs/2004.13721}{{\ttfamily arXiv:2004.13721
  [hep-th]}}.

\bibitem{Cheong:2023hrj}
D.~Y. Cheong, S.~C. Park, and C.~S. Shin, ``{Effective Theory Approach for
  Axion Wormholes},'' \href{http://arxiv.org/abs/2310.11260}{{\ttfamily
  arXiv:2310.11260 [hep-th]}}.

\bibitem{Loges:2023ypl}
G.~J. Loges, G.~Shiu, and T.~Van~Riet, ``{A 10d construction of Euclidean axion
  wormholes in flat and AdS space},''
  \href{http://dx.doi.org/10.1007/JHEP06(2023)079}{{\em JHEP} {\bfseries 06}
  (2023) 079}, \href{http://arxiv.org/abs/2302.03688}{{\ttfamily
  arXiv:2302.03688 [hep-th]}}.

\bibitem{Hertog:2017owm}
T.~Hertog, M.~Trigiante, and T.~Van~Riet, ``{Axion Wormholes in AdS
  Compactifications},'' \href{http://dx.doi.org/10.1007/JHEP06(2017)067}{{\em
  JHEP} {\bfseries 06} (2017) 067},
  \href{http://arxiv.org/abs/1702.04622}{{\ttfamily arXiv:1702.04622
  [hep-th]}}.

\bibitem{Astesiano:2022qba}
D.~Astesiano, D.~Ruggeri, M.~Trigiante, and T.~Van~Riet, ``{Instantons and (no)
  wormholes in $AdS_3\times S^3 \times CY_2$},''
  \href{http://arxiv.org/abs/2201.11694}{{\ttfamily arXiv:2201.11694
  [hep-th]}}.

\bibitem{Marolf:2021kjc}
D.~Marolf and J.~E. Santos, ``{AdS Euclidean wormholes},''
  \href{http://dx.doi.org/10.1088/1361-6382/ac2cb7}{{\em Class. Quant. Grav.}
  {\bfseries 38} no.~22, (2021) 224002},
  \href{http://arxiv.org/abs/2101.08875}{{\ttfamily arXiv:2101.08875
  [hep-th]}}.

\bibitem{Astesiano:2023iql}
D.~Astesiano and F.~F. Gautason, ``{Supersymmetric wormholes in String
  theory},'' \href{http://arxiv.org/abs/2309.02481}{{\ttfamily arXiv:2309.02481
  [hep-th]}}.

\bibitem{Anabalon:2023kcp}
A.~Anabal\'on, A.~Arboleya, and A.~Guarino, ``{Euclidean flows, solitons and
  wormholes in AdS from M-theory},''
  \href{http://arxiv.org/abs/2312.13955}{{\ttfamily arXiv:2312.13955
  [hep-th]}}.

\bibitem{Hamada:2019fmc}
Y.~Hamada, E.~Kiritsis, F.~Nitti, and L.~T. Witkowski, ``{Axion RG flows and
  the holographic dynamics of instanton densities},''
  \href{http://dx.doi.org/10.1088/1751-8121/ab4712}{{\em J. Phys. A} {\bfseries
  52} no.~45, (2019) 454003}, \href{http://arxiv.org/abs/1905.03663}{{\ttfamily
  arXiv:1905.03663 [hep-th]}}.

\bibitem{Maldacena:2004rf}
J.~M. Maldacena and L.~Maoz, ``{Wormholes in AdS},''
  \href{http://dx.doi.org/10.1088/1126-6708/2004/02/053}{{\em JHEP} {\bfseries
  02} (2004) 053}, \href{http://arxiv.org/abs/hep-th/0401024}{{\ttfamily
  arXiv:hep-th/0401024}}.

\bibitem{Arkani-Hamed:2007cpn}
N.~Arkani-Hamed, J.~Orgera, and J.~Polchinski, ``{Euclidean wormholes in string
  theory},'' \href{http://dx.doi.org/10.1088/1126-6708/2007/12/018}{{\em JHEP}
  {\bfseries 12} (2007) 018}, \href{http://arxiv.org/abs/0705.2768}{{\ttfamily
  arXiv:0705.2768 [hep-th]}}.

\bibitem{McNamara:2020uza}
J.~McNamara and C.~Vafa, ``{Baby Universes, Holography, and the Swampland},''
  \href{http://arxiv.org/abs/2004.06738}{{\ttfamily arXiv:2004.06738
  [hep-th]}}.

\bibitem{Katmadas:2018ksp}
S.~Katmadas, D.~Ruggeri, M.~Trigiante, and T.~Van~Riet, ``{The holographic dual
  to supergravity instantons in $\rm AdS_5\times S^5/\mathbb{Z}_k$},''
  \href{http://dx.doi.org/10.1007/JHEP10(2019)205}{{\em JHEP} {\bfseries 10}
  (2019) 205}, \href{http://arxiv.org/abs/1812.05986}{{\ttfamily
  arXiv:1812.05986 [hep-th]}}.

\bibitem{Gutperle:2002km}
M.~Gutperle and W.~Sabra, ``{Instantons and wormholes in Minkowski and (A)dS
  spaces},'' \href{http://dx.doi.org/10.1016/S0550-3213(02)00942-2}{{\em Nucl.
  Phys. B} {\bfseries 647} (2002) 344--356},
  \href{http://arxiv.org/abs/hep-th/0206153}{{\ttfamily arXiv:hep-th/0206153}}.

\bibitem{Aguilar-Gutierrez:2023ril}
S.~E. Aguilar-Gutierrez, T.~Hertog, R.~Tielemans, J.~P. van~der Schaar, and
  T.~Van~Riet, ``{Axion-de Sitter wormholes},''
  \href{http://arxiv.org/abs/2306.13951}{{\ttfamily arXiv:2306.13951
  [hep-th]}}.

\bibitem{Loges:2022nuw}
G.~J. Loges, G.~Shiu, and N.~Sudhir, ``{Complex Saddles and Euclidean Wormholes
  in the Lorentzian Path Integral},''
  \href{http://arxiv.org/abs/2203.01956}{{\ttfamily arXiv:2203.01956
  [hep-th]}}.

\bibitem{Coleman:1987rm}
S.~R. Coleman, ``{Quantum Tunneling and Negative Eigenvalues},''
  \href{http://dx.doi.org/10.1016/0550-3213(88)90308-2}{{\em Nucl. Phys. B}
  {\bfseries 298} (1988) 178--186}.

\bibitem{Kallosh:1995hi}
R.~Kallosh, A.~D. Linde, D.~A. Linde, and L.~Susskind, ``{Gravity and global
  symmetries},'' \href{http://dx.doi.org/10.1103/PhysRevD.52.912}{{\em Phys.
  Rev. D} {\bfseries 52} (1995) 912--935},
  \href{http://arxiv.org/abs/hep-th/9502069}{{\ttfamily arXiv:hep-th/9502069}}.

\bibitem{VanRiet:2020csu}
T.~Van~Riet, ``{A comment on no-force conditions for black holes and branes},''
  \href{http://dx.doi.org/10.1088/1361-6382/abe01a}{{\em Class. Quant. Grav.}
  {\bfseries 38} no.~7, (2021) 077001},
  \href{http://arxiv.org/abs/2010.11590}{{\ttfamily arXiv:2010.11590
  [hep-th]}}.

\bibitem{Bergshoeff:2004pg}
E.~Bergshoeff, A.~Collinucci, U.~Gran, D.~Roest, and S.~Vandoren,
  ``{Non-extremal instantons and wormholes in string theory},''
  \href{http://dx.doi.org/10.1002/prop.200410227}{{\em Fortsch. Phys.}
  {\bfseries 53} (2005) 990--996},
  \href{http://arxiv.org/abs/hep-th/0412183}{{\ttfamily arXiv:hep-th/0412183}}.

\bibitem{Breitenlohner:1987dg}
P.~Breitenlohner, D.~Maison, and G.~W. Gibbons, ``{Four-Dimensional Black Holes
  from Kaluza-Klein Theories},''
  \href{http://dx.doi.org/10.1007/BF01217967}{{\em Commun. Math. Phys.}
  {\bfseries 120} (1988) 295}.

\bibitem{Mukhanov:1992me}
V.~F. Mukhanov, H.~A. Feldman, and R.~H. Brandenberger, ``{Theory of
  cosmological perturbations. Part 1. Classical perturbations. Part 2. Quantum
  theory of perturbations. Part 3. Extensions},''
  \href{http://dx.doi.org/10.1016/0370-1573(92)90044-Z}{{\em Phys. Rept.}
  {\bfseries 215} (1992) 203--333}.

\bibitem{Gratton:1999ya}
S.~Gratton and N.~Turok, ``{Cosmological perturbations from the no boundary
  Euclidean path integral},''
  \href{http://dx.doi.org/10.1103/PhysRevD.60.123507}{{\em Phys. Rev. D}
  {\bfseries 60} (1999) 123507},
  \href{http://arxiv.org/abs/astro-ph/9902265}{{\ttfamily
  arXiv:astro-ph/9902265}}.

\bibitem{Pitrou:2013hga}
C.~Pitrou, X.~Roy, and O.~Umeh, ``{xPand: An algorithm for perturbing
  homogeneous cosmologies},''
  \href{http://dx.doi.org/10.1088/0264-9381/30/16/165002}{{\em Class. Quant.
  Grav.} {\bfseries 30} (2013) 165002},
  \href{http://arxiv.org/abs/1302.6174}{{\ttfamily arXiv:1302.6174
  [astro-ph.CO]}}.

\bibitem{Martin:2008xtensor}
J.~M. Mart{\'i}n-Garc{\'i}a, ``{xTensor: A free fast abstract tensor
  manipulator},'' in {\em The Eleventh Marcel Grossmann Meeting: On Recent
  Developments in Theoretical and Experimental General Relativity, Gravitation
  and Relativistic Field Theories (In 3 Volumes)}, pp.~1552--1554, World
  Scientific.
\newblock 2008.

\bibitem{Brizuela:2008ra}
D.~Brizuela, J.~M. Martin-Garcia, and G.~A. Mena~Marugan, ``{xPert: Computer
  algebra for metric perturbation theory},''
  \href{http://dx.doi.org/10.1007/s10714-009-0773-2}{{\em Gen. Rel. Grav.}
  {\bfseries 41} (2009) 2415--2431},
  \href{http://arxiv.org/abs/0807.0824}{{\ttfamily arXiv:0807.0824 [gr-qc]}}.

\bibitem{Maenaut:2020Git}
S.~Maenaut, ``{xPand} modifications for {Euclidean} spacetimes.''
  \url{https://github.com/SimonMaenaut/xPand}.

\bibitem{Garriga:1997wz}
J.~Garriga, X.~Montes, M.~Sasaki, and T.~Tanaka, ``{Canonical quantization of
  cosmological perturbations in the one-bubble open universe},''
  \href{http://dx.doi.org/10.1016/S0550-3213(97)00780-3}{{\em Nucl. Phys. B}
  {\bfseries 513} (1998) 343--374},
  \href{http://arxiv.org/abs/astro-ph/9706229}{{\ttfamily
  arXiv:astro-ph/9706229}}. [Erratum: Nucl.Phys.B 551, 511--511 (1999)].

\bibitem{Gratton:2001gw}
S.~Gratton, A.~Lewis, and N.~Turok, ``{Closed universes from cosmological
  instantons},'' \href{http://dx.doi.org/10.1103/PhysRevD.65.043513}{{\em Phys.
  Rev. D} {\bfseries 65} (2002) 043513},
  \href{http://arxiv.org/abs/astro-ph/0111012}{{\ttfamily
  arXiv:astro-ph/0111012}}.

\bibitem{Marolf:2020xie}
D.~Marolf and H.~Maxfield, ``{Transcending the ensemble: baby universes,
  spacetime wormholes, and the order and disorder of black hole information},''
  \href{http://dx.doi.org/10.1007/JHEP08(2020)044}{{\em JHEP} {\bfseries 08}
  (2020) 044}, \href{http://arxiv.org/abs/2002.08950}{{\ttfamily
  arXiv:2002.08950 [hep-th]}}.

\end{thebibliography}\endgroup

\end{document}